\documentclass[graybox,natbib]{svmult}

\usepackage{amssymb,amsmath}

\usepackage{mathptmx}       % selects Times Roman as basic font
\usepackage{helvet}         % selects Helvetica as sans-serif font
\usepackage{courier}        % selects Courier as typewriter font
\usepackage{type1cm}        % activate if the above 3 fonts are
\usepackage{makeidx}         % allows index generation
\usepackage{graphicx}        % standard LaTeX graphics tool
\usepackage{multicol}        % used for the two-column index
\usepackage[bottom]{footmisc}% places footnotes at page bottom

\usepackage{lipsum}
\usepackage{url}
\usepackage[section,ruled]{algorithm}
\usepackage{algorithmic}
\usepackage{boxedminipage}
\usepackage[bookmarks=true,linkcolor=blue,hyperfootnotes=false,breaklinks=true,citecolor=blue,colorlinks=true]{hyperref}
\usepackage{sistyle}
\SIthousandsep{,}

\makeindex             % used for the subject index
\begin{document}

\title*{Behavioral Modernity and the Cultural Transmission of Structured Information: The Semantic Axelrod Model}
\titlerunning{Cultural Transmission of Structured Information}
\author{Mark E. Madsen and Carl P. Lipo}
\institute{Mark E. Madsen \at Dept. of Anthropology, University of Washington, Box 353100, Seattle, WA 98195 \email{mark@madsenlab.org}
\and Carl P. Lipo \at Department of Anthropology and IIRMES, California State University at Long Beach, 1250 Bellflower Blvd, Long Beach, CA  90840 \email{Carl.Lipo@csulb.edu}}
\maketitle

\abstract*{Cultural transmission models are coming to the fore in explaining increases in the Paleolithic toolkit richness and diversity. During the later Paleolithic, technologies increase not only in terms of diversity but also in their complexity and interdependence. As \citet{Mesoudi2008a} have shown, selection broadly favors social learning of information that is hierarchical and structured.  We believe that teaching provides the necessary scaffolding for transmission of more complex cultural traits.  Here, we introduce an extension of the Axelrod \citeyearpar{axelrod1997} model of cultural differentiation in which traits have prerequisite relationships, and where social learning is dependent upon the ordering of those prerequisites. We examine the resulting structure of cultural repertoires as learning environments range from largely unstructured imitation, to structured teaching of necessary prerequisites, and we find that in combination with individual learning and innovation, high probabilities of teaching prerequisites leads to richer cultural repertoires.  Our results point to ways in which we can build more comprehensive explanations of the archaeological record of the Paleolithic as well as other cases of technological change.}

\abstract{Cultural transmission models are coming to the fore in explaining increases in the Paleolithic toolkit richness and diversity. During the later Paleolithic, technologies increase not only in terms of diversity but also in their complexity and interdependence. As \citet{Mesoudi2008a} have shown, selection broadly favors social learning of information that is hierarchical and structured.  We believe that teaching provides the necessary scaffolding for transmission of more complex cultural traits.  Here, we introduce an extension of the Axelrod \citeyearpar{axelrod1997} model of cultural differentiation in which traits have prerequisite relationships, and where social learning is dependent upon the ordering of those prerequisites. We examine the resulting structure of cultural repertoires as learning environments range from largely unstructured imitation, to structured teaching of necessary prerequisites, and we find that in combination with individual learning and innovation, high probabilities of teaching prerequisites leads to richer cultural repertoires.  Our results point to ways in which we can build more comprehensive explanations of the archaeological record of the Paleolithic as well as other cases of technological change.}

\section{Introduction}\label{introduction}

Although humans and our hominid ancestors have been cultural animals
throughout our evolutionary history, an important change occurred in our
lineage during the Middle and Upper Paleolithic. For millennia our
ancestors manufactured relatively small toolkits and their material
culture was remarkably similar across continental distances and over
many generations. Beginning in the Middle Paleolithic and continuing
through the Upper Paleolithic, the archaeological record reflects an
explosion in our cultural repertoire. Over tens of thousands of years,
artifactual toolkits shift from sets of relatively few objects with
multiple uses to large collections of functionally-specialized tools
that, employed increasingly complex technologies and that were
manufactured from an enriched range of materials. The changes in
artifacts suggest that human solutions to the problems of everyday life
became regionalized and differentiated. Further, the economic basis of
our lives began to broaden and also, in many areas, to become
specialized \citep{bar2002upper, d2011evolution, guy2005mosaic}.

While early researchers believed that the Upper Paleolithic resulted
from a singular ``revolution'' in human evolution leading to
behaviorally modern homo sapiens, this view is held by a minority of
paleoanthropologists and archaeologists today
\citep[e.g.,][]{klein2009human}. Careful examination of the Middle
Paleolithic archaeological record especially in Africa and the Near East
suggests that this change in behavior did not occur as a single distinct
event, instead occurring over a long period of time since much of the
enriched material culture we later characterize as the ``Upper
Paleolithic'' had precursors. In addition, this change now appears to be
patchy and fitful, with modern features appearing and frequently being
lost again
\citep{bouzouggar200782, d2007additional, d2011evolution, guy2005mosaic, mcbrearty2000revolution, mcbrearty2007down}.
Nor does behavioral modernity map neatly to biological taxa and their
movements, given that evidence for the precursors of fully modern
behavior is abundant in deposits associated with Neaderthals in addition
to modern \emph{Homo sapiens} \citep{Villa:2014kl}.

The ``learning hypothesis'' studied in this series of volumes makes the
plausible claim that behavioral modernity is the product of cumulative
changes in the way cultural information was acquired and retained across
generations \citep{Nishiaki2013Introduction}, thus providing a potential
explanation for the slow evolution of ``modern'' features, its
patchiness in space and time, and the lack of a neat mapping between
hominin taxonomy and material culture. In short, according to the
learning hypothesis, behavioral modernity arose through a change or
changes in the way social learning operated within hominin groups, with
those groups adopting richer modes of cultural learning surviving and
spreading compared to those who retained simpler forms of social
learning.

Within the umbrella of the learning hypothesis, there are many ways in
which social learning and thus intergenerational cultural transmission
could have changed, and an increasing amount of research is focused upon
formulating and testing different models. One class of studies is
focused upon factors exogeneous to the learning or imitation process
itself. Shennan
\citetext{\citeyear{shennan2000population}; \citeyear{shennan2001demography}}
proposed that population size has a powerful effect on diversity within
cultural transmission processes, which Henrich showed in the case of
toolkit element loss during a Tasmanian population bottleneck
\citep{henrich2004}. In a similar line of reasoning, Kuhn
\citeyearpar{Kuhn2013Cultural-Transm} argues that low population size
and density put Neanderthals in a situation where innovations spread
slowly and ultimately led to their demise relative to modern humans.
Furthermore, a growing set of experimental studies clearly show a
relationship between accumulation of complex cultural traits and the
number of cultural ``models'' from whom individuals can learn
\citep{muthukrishna2014sociality, derex2013experimental, kempe2014experimental}.
Not all studies have shown a strong association between population size
and cultural diversity, however. Collard and colleagues, find little
association in a linked series of comparative studies
\citep{collard2011drives, collard2013population, collard2013risk, collard2013plos}.
Finally, in his analysis of the overall evolutionary rate, Aoki
\citeyearpar{Aoki2013Determinants-of} found that innovation rates were
more important than population size to determining the rate of evolution
in a population.

To us, this body of work indicates that while population size is an
important parameter in mathematical models, it may be better understood
as a second-order effect in the real world, interacting with a myriad of
other factors and thus often dominated by those factors. Another
important factor is the structure of bands or demes into larger regional
metapopulations. Network topology, for example, is known to have a
substantial effect upon contagion or diffusion processes
\citep[e.g.,][]{castellano2009statistical, smilkov2012influence}. Thus,
it is likely that regional structure has critical effects on the
outcomes we can expect from a single social ``learning rule.'' Along
these lines, Premo \citeyearpar{premo2012local} has examined whether
metapopulation dynamics that include local extinction and recolonization
might provide an improved account for the retention and expansion of
diversity.

A second group of studies has focused upon endogeneous changes to social
learning processes. Many authors in this volume series, for example,
have looked at aspects of the way individuals learn skills and acquire
information \citep{Aoki2013Determinants-of, Nishiaki2013Introduction}.
We know that learning and teaching styles vary across human groups, and
formal modeling efforts are beginning to make clear that such variation
has evolutionary consequences that might lead to a rapid expansion of
the human cultural repertoire \citep{Nakahashi2013Cultural-Evolut}.
Those populations which increased the amount or effectiveness of
teaching would have a fitness advantage over those who relied upon
imitation and ``natural pedagogy'' in passing along technological and
foraging knowledge
\citep{Csibra:2011dx, Fogarty:2011gv, Terashima2013The-Evolutionar}.
Demography and population structure would then play an important role in
reinforcing the fitness differences which different learning strategies
would create, as pointed out by Kuhn
\citeyearpar{Kuhn2013Cultural-Transm}.

Ultimately, a full ``learning explanation'' for behavioral modernity
will be multifacted, including demographic and spatial changes as well
as changes to the mechanisms of social learning and technological
innovation themselves. Sterelny \citeyearpar[p.61]{sterelny2012evolved}
sums up this kind of multifactorial approach to behavioral modernity
well:

\begin{quote}
\ldots{}the cultural learning characteristic of the Upper Paleolithic
transition and later periods of human culture---social transmission with
both a large bandwidth and sufficient accuracy for incremental
improvement---requires individual cognitive adaptations for cultural
learning, highly structured learning environments, and population
structures that both buffer existing resources effectively and support
enough specialization to generate a supply of innovation.
\end{quote}

In research designed to explore how the structure of a learning
environment affects the results of social learning, Creanza and
colleagues \citeyearpar{Creanza2013Exploring-Cultu}, Aoki
\citeyearpar{Aoki2013Determinants-of}, Nakahashi
\citeyearpar{Nakahashi2013Cultural-Evolut}, and Castro and colleagues
\citeyearpar{Castro201474} developed models that examine how explicit
teaching (as opposed to simple imitation) affects the overall
evolutionary rate or cultural diversity in a population. Castro et al.,
for example, find that cumulative cultural transmission requires active
teaching in order to achieve fidelity across generations. Our work in
this chapter follows these authors, focusing on the nature of
transmitted information itself and the effects of teaching upon the
richness of structured technological knowledge.

In particular, we suggest that when knowledge is structured such that
skills and information must be learned in sequences, high fidelity
learning environments are critical to evolving ever-richer cultural
repertoires, of the type seen in behaviorally modern assemblages. To
formalize this idea, we construct a model which:

\begin{itemize}
\itemsep1pt\parskip0pt\parsep0pt
\item
  Represents cultural traits as hierarchically structured, in order to
  study increases in complexity,
\item
  Has a learning rule sensitive to the order in which cultural traits
  are acquired, with multiple levels of fidelity, and
\item
  Has a mechanism (such as homophily) that allows cultural
  differentiation endogeneous to the model.
\end{itemize}

As we alter the ``learning environment'' in our models from less to more
frequent teaching of traits and their prerequisites, we expect to see
greater diversity, larger structured sets of traits persisting in the
population, and greater differentiation of the population into
``different'' cultural configurations. We also expect that individual
innovation, independent of the social learning context, will play a role
in the accumulation of cultural complexity by allowing a population to
explore increasingly large spaces of technological design possibilities;
this expectation is concordant with Aoki's
\citeyearpar{Aoki2013Determinants-of} result in Volume I of this series.

In this chapter, we introduce a simulation model which combines a
hierarchical trait space capable of expressing dependencies or semantic
relationships between skills and information \citep{Mesoudi2008a}, and a
modified version of Robert Axelrod's \citeyearpar{axelrod1997}
homophilic social learning model which allows us to examine the
conditions under which evolution in a hierarchical design space leads to
cultural differentiation. After describing the model, we study its
dynamics and provide an initial assessment of its suitability for
studying the onset of behavioral modernity in the later Paleolithic.
Models like this begin to move beyond diffusion dynamics, bringing the
actual meaning and relations of traits into the modeling process. Hence,
we call these ``semantic Axelrod'' models, and believe that such models
form a platform for formalizing the type of multi-factor hypotheses
necessary to examine major transitions in human evolution, such as
``behavioral modernity.''

\section{The Semantic Axelrod Model for Trait
Prerequisites}\label{the-semantic-axelrod-model-for-trait-prerequisites}

Much of our technical knowledge, whether of stone tool manufacture,
throwing clay pots, or computer repair, is built from simple tasks, bits
of background knowledge, and step-by-step procedures
\citep{neff1992ceramics, schiffer1987theory}. These pieces of cultural
information are not simply a set of alternative options, which can be
mixed and matched in any combination. Instead, there are dependencies
and relationships between items which affect how skills and information
are learned and passed on between individuals. Some items will be
related in time, as steps in a process. Others will be related by
subsumption: arrowheads are a subclass of bifacial stone tools, and
require many of the same production techniques as bifaces used in other
projectiles. Still others will be related as sets of alternatives:
choices of surface treatment for a given ceramic paste, given the firing
regime selected, for example. To date, most archaeological models of
tool production have focused upon temporal relations in the construction
of an artifact, as in ``sequence models'' or
``$\textrm{cha\^ine op\'eratoire}$,'' but it is important to remember
that other representations are possible, including trees and more
general graphs to capture relations of use, reworking, or discard
\citep{Bamforth:2008kq, Bleed:2008in, Ferguson:2008ce, Hogberg:2008fj, bleed2001trees, bleed2002obviously, schiffer1987theory, stout2002skill}.

Given conscious reflection, we describe and organize our knowledge and
skills in many ways, but it is common (especially while learning a new
skill) to think of a complex process as a ``script'' or ``recipe''
\citep{schank1977scripts}. Experts in a task or field may not represent
their knowledge this way, having internalized such structures below the
conscious level. Experts will often know more than one way to accomplish
any given goal, and be able to repurpose and recombine methods and
tools, as opposed to the simpler, more linear or tree-based recipes of
the novice or student
\citep[e.g.,][]{Bleed:2008in, bleed2002obviously, stout2002skill}.
Nevertheless, it is common to teach or learn new information and skills
in a stepwise manner.

In this chapter, we focus not on the execution steps of a recipe (and
thus not on sequence models), but the relations between skills and
information \emph{during the learning process}. In specific, we focus
upon the \emph{prerequisite} relationships that exist between cultural
traits, since the ordered dependencies between skills and information
form one of the structures within social learning occurs during
development (and into adulthood). Some pieces of information or skills
must be in place before a person can effectively learn or practice
others. Examples from our own childhoods abound: one needed to
understand addition and subtraction and multiplication before learning
long division; in order to make soup, we need to understand how to
simmer rather than boil, how to chop and slice, what ingredients might
be combined, and so on. The fact that knowledge and skills build upon
one another make prerequisite relations between cultural traits
ubiquitous. In this chapter, we represent prerequisite relations as
trees in the graph-theoretic sense \citep{diestel2010graph}, replacing
the ``nominal scale'' structure of ``locus/allele'' models or
paradigmatic classifications and some typologies \citep{Dunnell1971},
but we emphasize that the tree models we discuss here are still
classifications and thus analytic tools, designed to allow us to measure
variation in the archaeological record, not reconstruct emic models of
Paleolithic technologies.

Our model also requires a way of representing a changing learning
environment, in ways that create higher fidelity and greater possibility
for building cumulative knowledge. In real learning environments, there
are many possibilities, but deliberate teaching and apprentice learning
are repeatedly seen across human groups as ways that naive individuals
can reliably learn the complex skills and information needed for
foraging, artifact production and maintenance, and navigating an
increasingly rich social world. The point of structuring the learning
environment with teaching and/or apprenticeship is to give the learner
skilled models to imitate, shortcut trial and error when acquiring a
skill, provide a reference for needed information, and to guide
individuals to put their information and skills together into
appropriate sequences to accomplish an overall goal. Apprenticeship and
formalized teaching provide a social learning ``scaffold,'' helping to
lower the amount of individual trial and error learning needed to master
a body of material \citep{wimsatt2007reproducing, wimsatt2007re}.

Within a standard discrete-time simulation model of a social learning
process, we can model this type of learning environment with the
following modifications:

\begin{enumerate}
\def\labelenumi{\arabic{enumi}.}
\itemsep1pt\parskip0pt\parsep0pt
\item
  Represent the order in which skills and information need to be
  acquired as a series of trees, with vertices representing traits
  (either a skill or piece of information), and edges the prerequisite
  relations between them.
\item
  Disallowing individuals the ability to copy traits from a cultural
  model for which they do not have necessary background or
  prerequisites, given the relations in the applicable tree model.
\item
  Creating a probability that individuals, if disallowed a trait, can be
  taught one of the needed prerequisites instead by that cultural model,
  leading to the potential accumulation of fuller knowledge and skills
  over time.
\end{enumerate}

By changing the probability that individuals learn a missing
prerequisite trait, we can ``tune'' the learning environment. Low
probabilities might correspond, for example, to a learning environment
where individuals can observe others executing a production step, but
are given little or no instruction or guidance on what they need to know
in order to successfully master it. High probabilities of learning
prerequisites would correspond, on the other hand, to environments where
individuals receive instruction, or work together with a more skilled
individual who guides them toward learning the information and skills
they lack. In the next section, we discuss our model of trait
relationships and the learning environment in more detail.

\subsection{Representation of Traits And Their
Prerequisites}\label{representation-of-traits-and-their-prerequisites}

In order to represent the ``prerequisite'' relations between a number of
cultural traits, we organize the traits into trees\footnote{A tree is a
  graph with no cycles or loops. That is, a tree is a connected graph on
  $n$ vertices that possesses at most $n-1$ edges
  \citep{diestel2010graph}. Furthermore, in this chapter we are
  concerned with \emph{rooted} trees, in which one vertex is
  distinguished as the ``origin'' of the tree, giving rise to a
  hierarchical structure.}, where nodes higher in the tree represent
knowledge, skills, or concepts which are necessary for traits further
down the tree. Let us consider the different skills and information
necessary for the construction of a single artifact, say a dart thrown
by an atlatl. An artisan will possess information about different raw
materials, an understanding of what materials are suitable for specific
purposes, skills and information concerning the knapping of different
types of bifaces, methods of hafting bifaces into different kinds of
shafts, and so on. Stout \citeyearpar{stout2011stone} organized such
knowledge into ``action hierarchies,'' which represent sequences of
actions, sets of choices, and optional elements for the construction of
a class of stone tools, drawing the representation from Moore's
\citeyearpar{moore2010grammars} graphical notation.

\begin{figure}[h] 
\centering 
\includegraphics[]{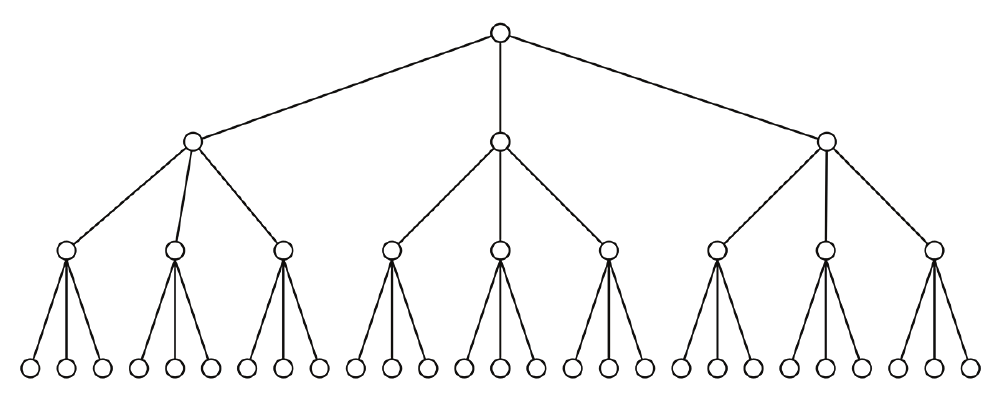} 
\caption{A single trait tree, represented by a balanced tree with branching factor 3 and depth factor 3, order 40.  In our model, nodes higher in the tree represent prerequisites for nodes lower down the tree.  Each instance of the model will have several or many of these trees in the design space.} 
\label{img:trait-tree} 
\end{figure}

We should emphasize that employing tree structures to represent learning
dependencies is a modeling choice. Other choices may be sensible as
well. General graphs could represent webs of relations between concepts
or skills, and multigraphs (replacing adjacency matrices with tensors)
can represent different types of relations in a single structure
\citep{ICML2011Nickel_438}. For purposes of the present chapter, we are
interested in the order in which people usually \emph{learn} skills and
information, rather than the order in which steps are executed. The
difference is potentially significant, in that two adjacent steps in a
sequence might involve very different information, tools, or skills,
which can be learned in parallel without dependencies. Because, in our
model, traits cannot be learned unless an individual possesses the
necessary prerequisites, we introduce the idea of a ``learning
hierarchy,'' which is a division of Stout's action hierarchy into
components which are learned with ordered dependencies, and independent
components represented in separate trees. For example, one might learn
about the sources of good lithic raw materials, independent of learning
how to perform different percussion techniques. In our model, each of
these independent areas is represented by a separate tree of traits.

\begin{figure}[h] 
\centering 
\includegraphics[scale=0.6]{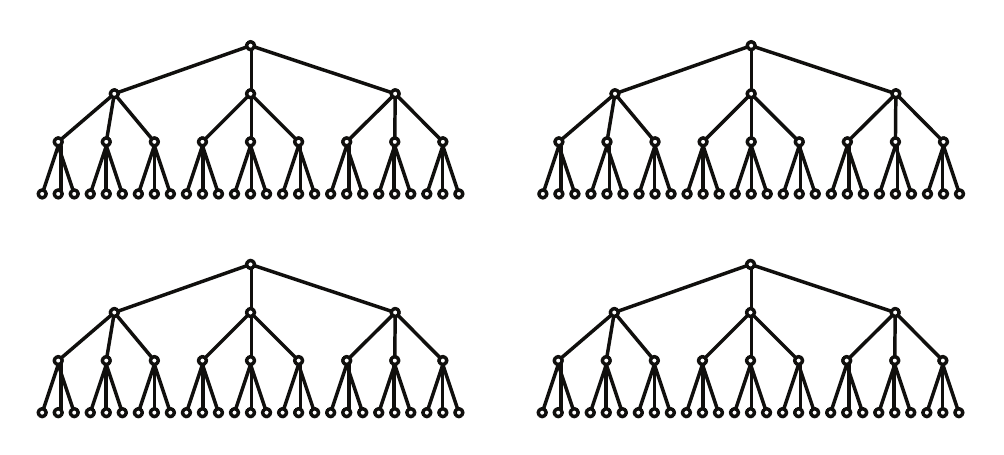} 
\caption{A design space composed of 4 independent trees, each tree with branching factor 3 and depth factor 3, order 40.  We also studied larger design spaces with 16 independent trees, and with larger branching and depth factors.} 
\label{img:design-space-4} 
\end{figure}

In each simulation model, we begin with a trait or ``design space'' that
incorporates several independent sets of traits \citep{o2010cultural}.
The overall design space of a simulation model is thus a
forest\footnote{A forest is a graph composed of multiple components,
  each of which is a tree.}, composed of several trees (Figure
\ref{img:design-space-4}). For each tree in a learning hierarchy, we
employ balanced trees which have the same number of nodes at each level,
to provide a simplified model of a design space with which to begin our
exploration of this class of social learning model, although real design
spaces are undoubtedly more complex in their geometry. Each tree in our
model is specified by a branching factor $r$ and depth $h$. As a result,
each trait tree in the design space has $\sum_{i=0}^{h} r^i$ traits.

The tree depicted in Figure \ref{img:trait-tree} thus has 40 vertices,
for example. In this chapter, we examine both small (4 trees) and larger
(16 trees) design spaces, to see how learning may differ in problems
involving design spaces of different size and complexity. We examine
trees with combinations of branching and depth factors of 3 and 5. Thus,
a design space with 4 trees with branching and depth factors of 3 (as in
Figure \ref{img:trait-tree}) would have 160 traits, whereas a design
space with 16 trees of branching and depth factors of 5 would have a
total of 62,496 traits.\footnote{We initially chose 6 as the limit on
  branching and depth factors, but found that we cannot calculate
  certain symmetry statistics, such as the size of the automorphism
  group, on trees that large using existing tools. Even a tree with
  $r=5, h=6$ has over $10^{1623}$ possible symmetries, and an attempt to
  calculate the symmetries for $r=6, h=6$ did not complete given the
  memory limits of the computers we had available.} Even the small
design spaces we consider here create a large space for cultural change
and differentiation, given the number of possible trees one can
construct on even 40 vertices.\footnote{If we consider each trait to be
  unique and non-interchangeable, the number of unique trees with unique
  vertex labels is $n^{n-2}$ by Cayley's theorem
  \citep{diestel2010graph}. For example, for each trait tree of 40
  vertices, there are roughly $10^{60}$ possible trees. Even if we
  consider traits to be interchangeable (e.g., we look at the abstract
  topology of trees rather than the details of individual traits), there
  are \emph{at least} $10^{16}$ possible unlabelled rooted trees on 40
  vertices (using Otter's \citeyearpar{otter1948number} approximation).}
In the experiments reported here, the overall size of the design space
remains constant over time, which is a simplifying assumption as we
develop this class of structured information models. In future work, we
will explore the role of invention in episodically creating large new
regions of design space for the evolving population to explore.

Given the total ``design space'' represented by a forest of trait trees,
each individual in our model is initialized with a small number of
``initial'' traits. Initial traits are chosen randomly but heavily
weighted towards the roots of the trees to represent the fact that our
knowledge starts out basic and sparse. In general, all of the design
spaces modeled here are larger than populations will explore within the
bounds of a simulation run. In the next sections we describe the social
learning model, modified from Robert Axelrod's original, by which each
simulated population evolves within this tree-structured design space,
and will return to the specifics of how an initial culture repertoire is
chosen.

\subsection{The Axelrod Model of Social Learning and
Differentiation}\label{the-axelrod-model-of-social-learning-and-differentiation}

Robert Axelrod \citeyearpar{axelrod1997} formulated a model aimed at
studying the conditions under which simple learning rules could lead to
cultural differentiation, rather than a single fixed state (which is the
result of simpler neutral or diffusion models). This makes it useful as
a starting point for understanding phenomena such as behavioral
modernity, in our view. Axelrod's model combines social learning, in the
form of random copying, a spatial structure to interaction, in the form
of localized copying of neighbors on a lattice, and the tendency to
interact most strongly with those to whom we are already culturally
similar (homophily). The model displays a rich and interesting set of
behaviors, and has been extensively studied by social scientists and
physicists \citep{castellano2009statistical}. First we review the basic
model, and in the following section our modified algorithm.

\subsubsection{Axelrod's Original Model}\label{axelrods-original-model}

The original model locates $N$ individuals on the nodes of a regular
lattice or grid, but various network structures have also been studied.
Each individual is endowed with $F$ integer variables
$(\sigma_1,\ldots,\sigma_F)$, that can each assume $q$ values. In the
original model, each variable is a ``cultural feature'' each of which
can assume $q$ ``traits.'' In each step, a randomly chosen individual
$i$ and a random neighbor $j$ are selected, and ``interact'' with
probability equal to the overlap between their cultural repertoire.
Overlap, in the basic model, is simply the fraction of features for
which $i$ and $j$ possess the same trait value:

\begin{equation}\label{eq:axelrod}p(i,j) = \frac{1}{F} \sum_{f=1}^{F} \delta_{\sigma_f(i)\sigma_f(j)}\end{equation}

where $\delta_{i,j}$ is Kronecker's delta function, taking the value $1$
when its two arguments are equal and $0$ otherwise. When individuals
interact, the focal individual $i$ takes the trait value of its neighbor
for one of the features where the two individuals differ.

Interaction has no effect when two individuals already possess identical
cultural repertoires, and there is no probability of interaction if
individuals have no traits in common. This eventually causes the model
to reach an absorbing state where no further changes are possible.
Instances of the model are initialized with a random distribution of
traits among individuals, and left to update until the steady state is
reached. The evolution of the population leads to two classes of
absorbing states: (a) a ``monocultural'' state in which all individuals
share the same set of variables, and (b) a ``polycultural'' state in
which subpopulations exist which share the same set of variables within
the group, and are completely different from their neighbors.

Which of the two results is reached, and the statistical character of
``polycultural'' states when they exist, depends mainly upon the number
of traits possible $q$ for each cultural feature. For small values of
$q$, individuals share many traits with their neighbors, interactions
are thus frequent, and one domain comprising a single set of traits will
grow to become fixed within the entire population. In contrast, when the
value of $q$ is high, individuals start out sharing very few traits,
with interactions that are correspondingly less frequent. Regions of
uniform cultural variation do grow, but as they do, sets of individuals
who share no traits at all (and thus do not interaction) grow as well,
and often prevent any single regional culture from expanding to fix
within the population.

Many variants of the basic Axelrod model have been studied, including
the addition of ``drift'' via the introduction of copying error,
situating agents on different types of complex networks, the addition of
an external ``field'' to simulate the effects of mass media, and copying
that obeys a ``conformist'' or majoritarian rule by selecting the most
common trait among the neighbor set
\citep{castellano2000nonequilibrium, de2009effects, flache2006sustains, GonzalezAvella:2007p6910, GonzalezAvella:2007p6912, gonzalez2005nonequilibrium, gonzalez2006local, Klemm:2003p7031, Klemm:2003p7112, Klemm:2005tb, Lanchier:2010p16999, Lanchier:2012ur}.
In general, modifications of the basic model can reduce the tendency of
the model to produce polycultural solutions, or change the time scale or
location of the critical point.

\subsubsection{Semantic Extensions to the Axelrod
Model}\label{semantic-extensions-to-the-axelrod-model}

\begin{figure}[htbp] 
\centering 
\includegraphics[scale=0.5]{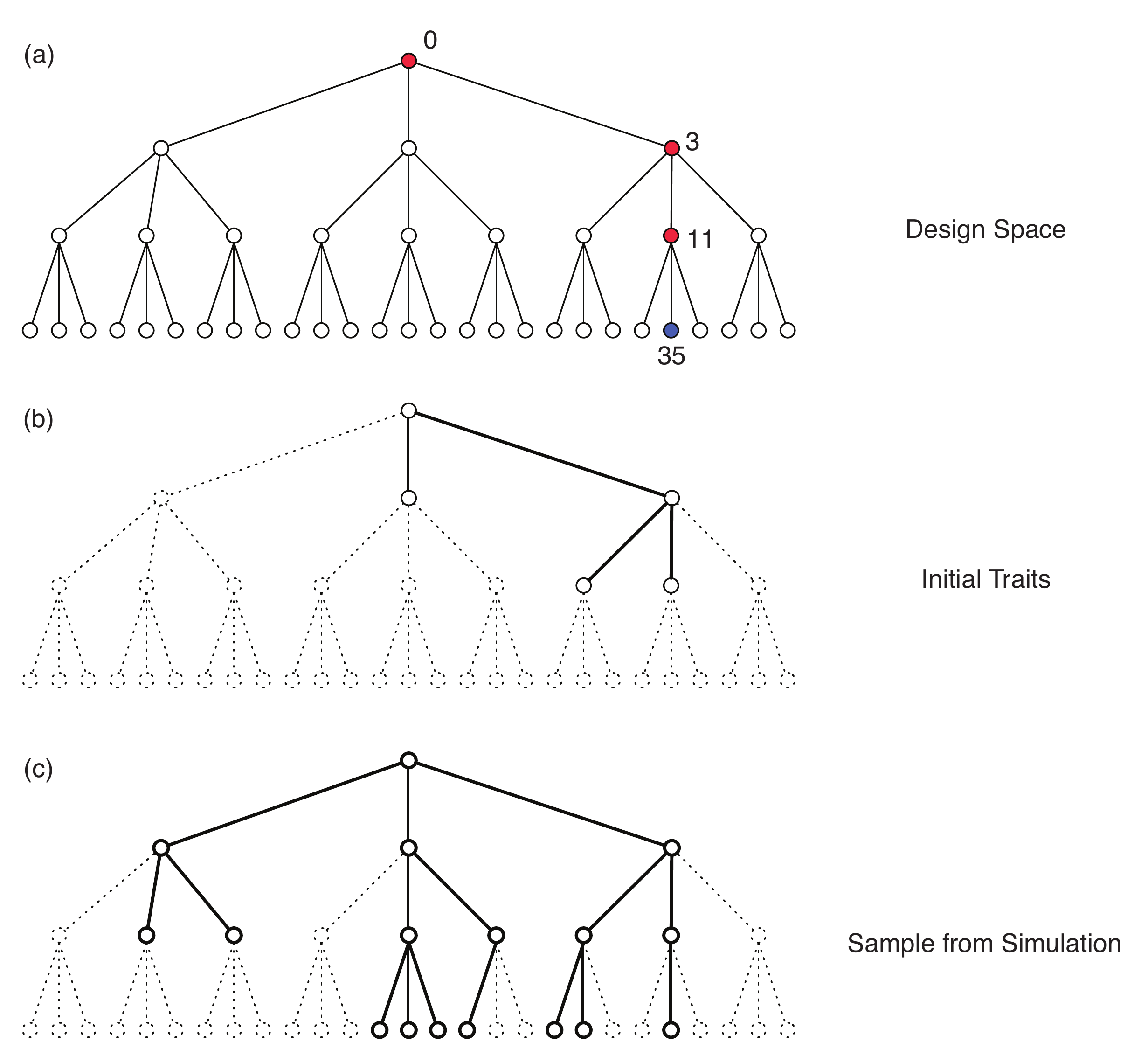} 
\caption{Illustration of a design space composed of a single trait tree, along with a random initial trait chosen from the design space, and a final sample from a simulation run, showing the evolution of traits within the design space.  Also shown in the top panel are the ``prerequisites'' for a cultural trait (35), as an example.} 
\label{img:prereq} 
\end{figure}

We begin each simulation with $N$ (100, 225, or 400) agents, arranged on
a square grid. A design space is created, with some number of trait
trees (4 or 16), with uniform branching factors and depth factors (3 or
5). An example of such a tree is shown in panel A of Figure
\ref{img:prereq}. Initial traits (and their prerequisites) are chosen
randomly across the configured number of trait trees, as follows. For
each individual, we select a random number $t$ between 1 and 4, and
repeat the trait selection process $t$ times for that individual. In
each selection, we choose a random tree in the design space, and then
select a depth in the tree for the trait, given by
$d  \sim \textrm{Poisson}(0.5)$. This biases trait selection towards the
root of the tree, as one would expect in young or inexperienced
individuals. We then walk $d$ steps into the tree, making uniform random
selections for the children of each vertex. The path of vertices thus
constructed is added to the individual's trait set, giving them an
initial trait and its necessary prerequisites. One such initial trait is
shown in Panel B of Figure \ref{img:prereq}. Given that individuals
begin with a small number of initial traits (between 1 and 4, selected
randomly), and their prerequisites, the initial trait endowment of an
individual is between 1 and $4h$, where $h$ is the maximum depth of the
design space (either 3 or 5 in the experiments reported here).\footnote{At
  maximum, this yields some individuals who begin the simulation with up
  to 20 traits. The median number of traits in samples taken after 6-10
  million time steps is considerably higher--259 traits per cultural
  configuration or region. Thus, cultural repertoires in the simulation
  grow through copying and innovation, as expected.}

Once the population is initialized, the simulation runs a discrete
approximation to a continuous-time model. In other words, only one agent
changes at each elemental time step, as in the original Axelrod model
and the Moran model of population genetics and its cultural version
\citep{aoki2011rates, moran1962statistical, moran1958random}. At each
step, an agent (A) is chosen at random, and a random neighbor of A is
then selected (agent B). Their probability of interaction is given by
the overlap of trait sets, which is most simply calculated as the
Jaccard overlap between the set of tree vertices each possesses, thus
replacing Equation \ref{eq:axelrod} with:

\begin{equation}J(A,B) = \frac{|V(A) \cap V(B)|}{|V(A) \cup V(B)|}\end{equation}

where $V(i)$ represents the vertex list for trait trees held by
individual $i$ in the population.

If the agents end up interacting, agent A observes the traits currently
possessed by B, and selects a trait (T) that it does not already possess
to learn. If agent A has the necessary prerequisite traits for the
selected trait, it can learn trait T. If not, there is a probability
$\mathbb{P}(l)$ that B can teach A a necessary prerequisite for T
instead. This simulates the process of agent B structuring the learning
environment of A through formal instruction or apprenticeship, for
example. If such a prerequisite learning event occurs given
$\mathbb{P}(l)$, agent A learns the most fundamental of T's
prerequisites that it does not already possess. For example, agent A
might require the trait closest to T (e.g., trait 11 in Figure
\ref{img:prereq}, if the original trait targeted was 35).

Additionally, at each time step, there is a probability $\mathbb{P}(m)$
that one random individual in the population will learn a new trait (and
necessary prerequisites) that it does not already possess. For example,
if an innovation event occurs and an agent discovers trait 35 by
individual trial and error learning, we assume that the agent also
discovered traits 0, 3, and 11. Thus innovation can introduce one trait
to the population, or a linked set depending upon its prerequisites and
what the innovating individual already ``knows.'' This model of
innovation simulates an ongoing process of individual learning
unconnected to social learning or teaching within the population.
Because this functions much like ``infinite-alleles mutation'' in the
classical Wright-Fisher neutral models \citep{Ewens2004}, or like noise
terms in Axelrod, Ising, or Potts models
\citep{castellano2009statistical}, we will refer to this as the ``global
innovation rate'' in this chapter.

Each simulation run lasts $10^7$ steps, which yields between $10^4$ and
$10^5$ copying events per individual, depending upon population
size.\footnote{100,000 was chosen as a compromise for running large
  batches of simulations in parallel. Some simulation runs, especially
  in small design spaces with very high prerequisite learning rates, can
  converge to a monocultural solution and quasi-stable equilibrium quite
  quickly; in the largest design spaces and low learning rates,
  convergence may never occur even though the process is well-mixed.
  However, the processes have reached a quasi-stable equilibrium,
  verified by examining samples at different times for secular trends in
  median and mean values, which were not found.} Samples are taken
beginning at 6 million steps, and sampling at an interval of 1 million
steps, and record the trait trees seen in the population. An example of
such a sampled tree is shown in Panel C of Figure \ref{img:prereq}. For
reference, the full algorithm for each copying step is given in the
Appendix as Algorithm \ref{alg:tree-prereq-axelrod}.

\section{Measuring Cultural Diversity and the Results of Structured
Learning}\label{measuring-cultural-diversity-and-the-results-of-structured-learning}

Each sample from a simulation run is composed of the distinct sets of
trait trees possessed by individuals in the population, along with
summary statistics. If a simulation run converges to a monocultural
solution, the sample will have one set of trait trees, which are shared
across the entire population. In other cases, there will be clusters of
cultural configurations which might be unique to a single individual, or
shared by some number of agents. Each cluster will be composed of some
number of trait trees (typically, the number configured for the
simulation run: 4 or 16, but perhaps a subset), and each trait tree will
be the result of many agents learning traits and their prerequisites
socially, and for runs with a non-zero mutation rate, by individual
learning or innovation. Each cluster will thus have some number of
traits, typically higher (often much higher) than the initial endowment
given to the population.

\begin{figure}[htbp] 
\centering 
\includegraphics[]{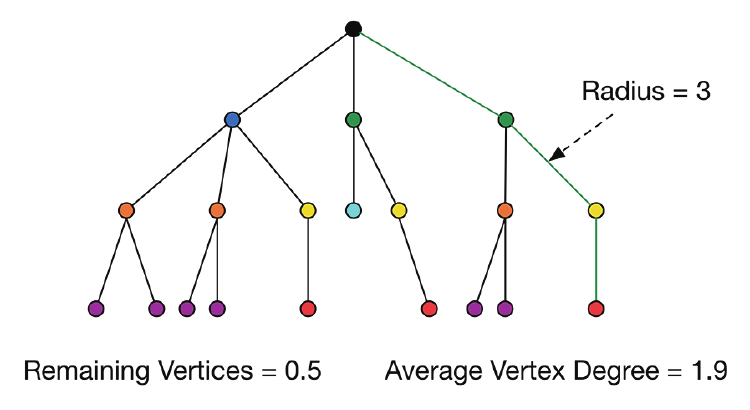} 
\caption{An example set of traits at the conclusion of a simulation run, extracted from a simulation with branching factor 3 and depth factor 3, and a single trait tree as the trait space.  The remaining density of vertices, mean vertex degree, and radius of the tree are noted.  Vertex colors denote ``structural equivalence'' classes or ``orbit structure,'' as measured by adjacency patterns, and is one measure of the symmetries present in the tree.} 
\label{img:final-tree} 
\end{figure}

From the sampled trait trees, we calculate summary statistics as
follows. The ratio of the number of traits in the sample to the full
design space size (or ``remaining density'' of traits) is one measure of
trait richness. The radius of a rooted tree is the number of edges in
the path from root to the furthest edge. The average radius of trees in
a sample (or its ratio to the depth of the design space) is another
richness measure, aimed at measuring whether knowledge with multiple
prerequisites is being learned within the simulated population.
Similarly, in the original design space, the branching factor describes
how many children each node in the tree started with, so measuring the
average vertex degree gives us a rough measure of how broad a cultural
repertoire is. Each of these measures is illustrated in Figure
\ref{img:final-tree} for an example tree selected from our data.

In addition to these simple numerical measures comparing final trees to
the original design space, it is useful to measure something about the
overall ``shape'' of the trees themselves. One way of formalizing this
notion is to examine the \emph{symmetries} of the final trait trees.
Examining Figure \ref{img:final-tree}, if we ignore the exact identities
of traits for the moment, it is apparent that there are repeating
patterns. For example, the left-most branches each terminate in a pair
of leaves. This pattern is repated on the second right-most branch.
These types of repeating patterns are computationally expensive to
search for in large sets of trees, but we can summarize them by
considering trait trees as algebraic objects and examining their
\emph{automorphisms}.

An automorphism is a function which maps an object to itself, in such a
way that the structure of the object is preserved
\citep{rotman1995introduction}. Graph automorphisms map vertices in a
graph to each other, preserving properties such as the adjacency pattern
of edges. The six vertices which mark the repeating pattern of
leaf-pairs in Figure \ref{img:final-tree} are an automorphism of the
tree, and thus are symmetries we can measure. An overall measure of
``how symmetrical'' (or ``how many interchangeable patterns'') there are
in a graph possesses given by the total number of automorphisms found,
called the size of the automorphism group or $|\textit{Aut}(G)|$
\citep{godsil2001algebraic}. A tree with no repeating patterns will thus
have an automorphism group size of 1, indicating that the only symmetry
is the entire tree itself. A balanced tree with branching and depth
factors of 3, as depicted in Figure \ref{img:trait-tree}, has
approximately $1.3 \times 10^{10}$ automorphisms. The more repeating
patterns there are in trait trees, the more automorphisms they will
possess.

Because group sizes grow quickly and the accuracy of performing
calculations with truly astronomical numbers is low, another possible
measure of the symmetries present is to count the \emph{classes} of
equivalences into which vertices fall. The \emph{orbits} of the
automorphism group are the sets of vertices which are interchangable by
some permutation that preserves structure. For example, the graph in
Figure \ref{img:trait-tree} has five orbits, with each vertex at a given
level interchangable (in a structural sense). Similarly, the six leaf
vertices that are part of pairs in Figure \ref{img:final-tree} are part
of the same orbit; in this illustration, each orbit is given a different
color to highlight their equivalence. For each cultural region found
when sampling a simulation, we calculate the size of the automorphism
group and the number and multiplicity (frequency) of orbits. For this
analysis, we employ the \emph{nauty + Traces} software by Brendan McKay
and Adolfo Piperno \citep{McKay201494}.\footnote{Nauty+Traces can be
  downloaded at \url{http://pallini.di.uniroma1.it/}. We employed
  version 2.5r7 for this research.}

\section{Experiments}\label{experiments}

Given a modified Axelrod model on a tree-structured trait space, we
expect to see greater cultural diversity, differentiation among groups
of individuals, and larger sets of traits as the ``learning
environment'' is tuned from a low to high probability of teaching and
learning among individuals. We also expect that individual innovation,
independent of the social learning context, will increase the amount of
the technological design space that a population explores, which leads
to enhanced opportunities for differentiation even through simple random
copying. Here we measure cultural differentiation by the number of
clusters of individuals who share the same trait trees when we sample
the population.

Second, we looked at whether highly structured learning environments,
represented here by higher probabilities of naive individuals gaining
the prerequisites for the skills and information they encounter with
peers, led to deeper and richer cultural repertoires. We explore a
number of ways of measure the richness of a cultural repertoire in a
model with structured relations between traits, through the use of graph
properties and symmetry measures. The measures used are those described
above: the tree radius (or depth), mean vertex degree, the fraction of
remaining vertices, and the size of the automorphism group of sampled
trait forests. Finally, we began to examine how the structured learning
environment might interact with demography, by simulating the same
parameters across two sizes of population.

For this chapter, we examined populations of size 100, 225 and 400, to
begin to examine the effects of population size. For these populations,
we examined design spaces that were small (4 trait trees) and large (16
trait trees). Within each size, we further examined combinations of
branching factor and depth factor with values of 3 and 5, thus yielding
8 total sizes of design space (Table
\ref{tab:axelrod-design-space-size}).

\begin{table}[H]
\begin{tabular}{|c|c|c|c|}
\hline
\textbf{Branching Factor} & \textbf{Depth Factor} & \textbf{Number of Trait Trees} & \textbf{Size of Design Space}\\ 
\hline
3 & 3 & 4 & 160\\ 
\hline 
5 & 3 & 4 & 624\\ 
\hline 
3 & 5 & 4 & 1456\\ 
\hline 
5 & 5 & 4 & 15624\\ 
\hline 
3 & 3 & 16 & 640\\ 
\hline 
5 & 3 & 16 & 2496\\ 
\hline 
3 & 5 & 16 & 5824\\ 
\hline 
5 & 5 & 16 & 62496\\ 
\hline 
\hline
\end{tabular}
\caption{Size of design space for different trait tree configurations}
\label{tab:axelrod-design-space-size}
\end{table}

Further, we examined three levels of global mutation or innovation rate:
zero, or no mutation, and 0.00005 and 0.0001. Such rates created a
constant supply of new innovations, but several orders of magnitude less
frequent than copying and prerequisite learning events. The full set of
parameters are given in Table \ref{tab:axelrodct-sim-parameters}. In
this pilot study, for each combination of all of the above parameters,
we performed 25 replications. With 5 samples per simulation run, this
yielded 10,963,691 samples of cultural regions.

\begin{table}[H]
\begin{tabular}{|p{0.6\textwidth}|p{0.4\textwidth}|}
\hline
\textbf{Simulation Parameter} & \textbf{Value or Values} \\ 
\hline
Population rate at which new traits arise by individual learning & 0.0, 5e-05, 0.0001\\ 
 \hline 
Maximum number of initial traits (not including their prerequisites) each individual is endowed with & 4\\ 
 \hline 
Number of distinct trees of traits and prerequisites & 4, 16\\ 
 \hline 
Population sizes & 100, 225, 400\\ 
 \hline 
Replicate simulation runs at each parameter combination & 25\\ 
 \hline 
Maximum time after which a simulation is sampled and terminated & 10000000\\ 
 \hline 
Individual probability for being taught a missing prerequisite & 0.05, 0.1, 0.2, 0.3, 0.4, 0.5, 0.6, 0.7, 0.8, 0.9\\ 
 \hline 
Number of branches at each level of a trait tree & 3, 5\\ 
 \hline 
Depth of traits in each trait tree & 3, 5\\ 
 \hline 
\hline
\end{tabular}
\caption{Parameter space for simulations described in this chapter}
\label{tab:axelrodct-sim-parameters}
\end{table}

\section{Results}\label{results}

We begin by noting that compared to the original Axelrod model, or
neutral and biased copying models, the dynamics of our semantic Axelrod
model are highly variable. A very wide range of outcomes is possible for
each parameter combination, especially when the size of the design space
is large. Some variables, such as the average vertex degree of sampled
trait trees, are strongly overlapping across all learning rates and do
not appear diagnostic of different learning environments, at least in
these initial experiments. Given the large amount of variability in the
dynamics, larger numbers of replications would be useful, although this
is computationally quite expensive at present.\footnote{The simulations
  reported here ran on a cluster of 6 compute-optimized ``extra large''
  Linux instances on Amazon's EC2 computing cloud, for a total of 17
  days of wall clock time and 2075 CPU hours. We plan further
  optimizations to the simulation code to make larger samples
  economically feasible.} That said, several features of the data are
strongly suggestive that hierarchical trait models have potential in
modeling cumulative technological evolution, making the computational
expense worthwhile.

\subsection{Cultural Diversity}\label{cultural-diversity}

\begin{figure}[htbp]
    \centering
    \includegraphics[scale=0.4]{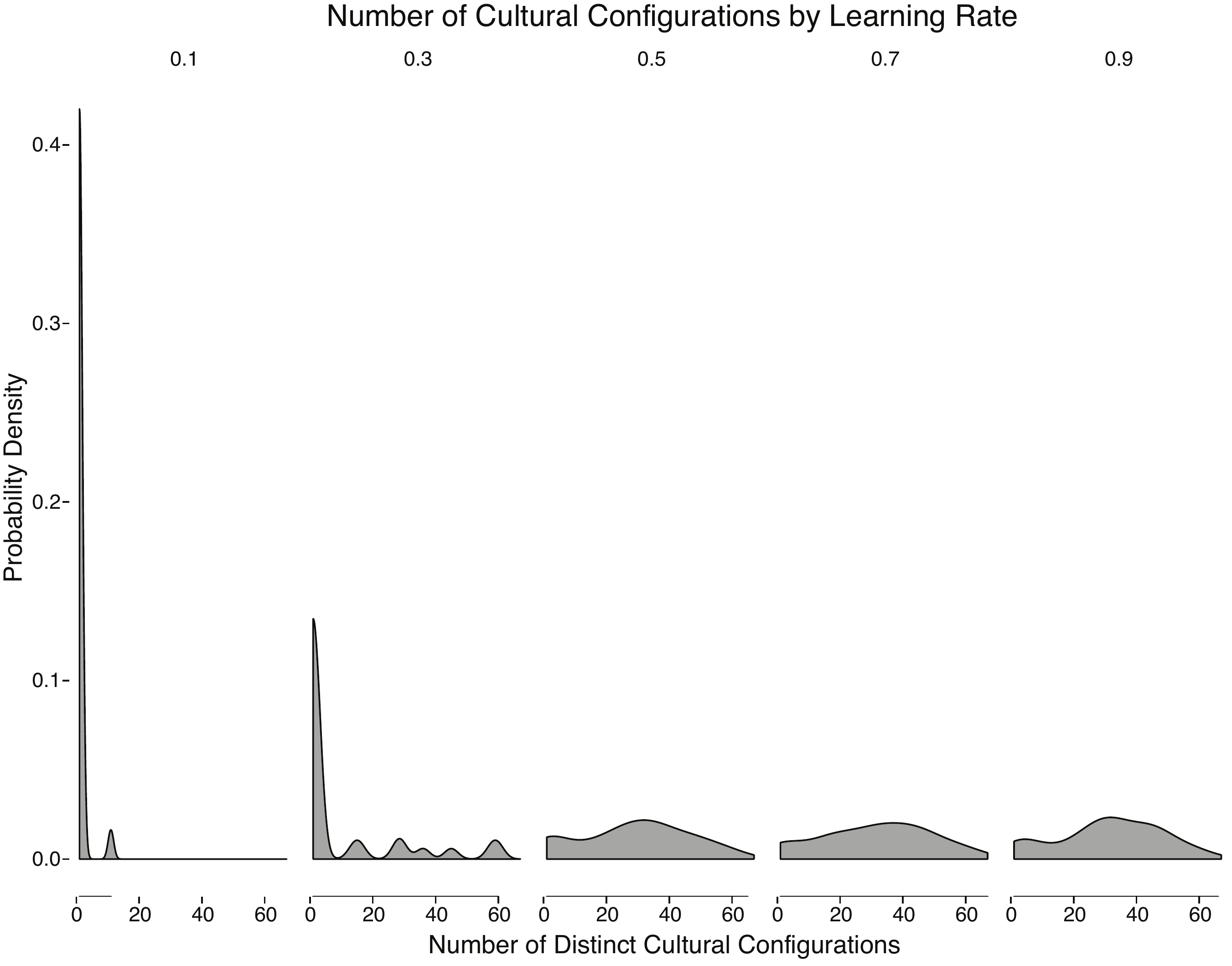}
    \caption{Number of cultural configurations in simulations with the smallest trait space (160 total traits in 4 trees), and a high individual innovation rate ($10^{-4}$).}
    \label{img:culture-count-sm-trait-high-innov}
\end{figure}

Variation among individuals is foundational to evolutionary processes,
and is the raw material from which differentiation between regions and
cultural groups is constructed. Figure
\ref{img:culture-count-sm-trait-high-innov} depicts the number of
cultural configurations (i.e., trait trees) in a population of size 100,
for the smallest trait space with only 160 total traits, and relatively
high levels of individual innovation. For example, in the left-most
panel the large peak just above zero indicates that most simulated
populations are characterized by one or a few sets of trait trees. Five
learning rates are depicted, increasing from left to right across the
panels. At the very lowest rate of learning fidelity, with only a 10\%
chance of being taught a needed prerequisite for knowledge being copied,
most of the populations simulated share a single set of traits, and even
individual innovation does not drive significant exploration of the
space of structured traits. With increased fidelity in teaching needed
prerequisites, however, simulated populations begin exhibiting marked
differentiation, with individuals possessing more unique configurations
of traits from the overall design space.

\begin{figure}[htbp] 
    \centering
    \includegraphics[scale=0.4]{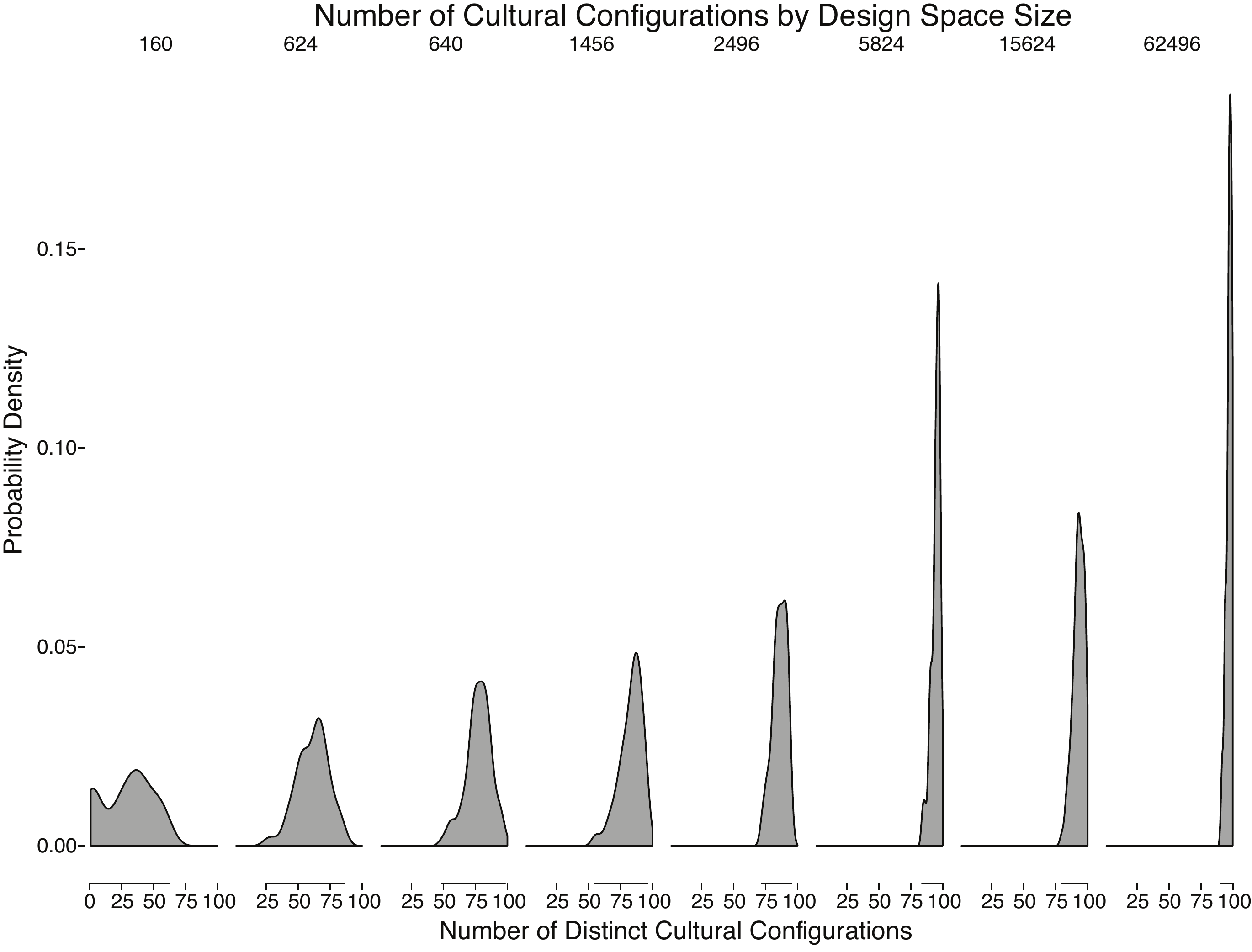}
    \caption{Number of cultural configurations in simulations with an intermediate learning rate (0.4), across different sizes of trait space.}
    \label{img:culture-count-lr04-by-traitspace-size}
\end{figure}

Looking at the data from another perspective, we can hold the fidelity
of learning constant (say, at a 40\% chance of being taught a needed
prerequisite), with the same global innovation rate ($10^{-4}$) as
Figure \ref{img:culture-count-sm-trait-high-innov}, and examine the
effect of different size design spaces (Figure
\ref{img:culture-count-lr04-by-traitspace-size}). In general,
populations exhibit greater differentiation between individuals as the
design space gets larger, as prerequisite learning helps individuals
acquire adjacent traits, and individual innovation randomly explores
more distant portions of the design space.

Given the structure of the Axelrod model, with the strong tendency
towards cultural uniformity given homophily, all simulated populations
converged to a single cultural configuration in the absence of a global
innovation rate. This highlights the importance of various
``innovation'' and ``invention'' processes in the creation and
maintenance of cultural differentiation and diversity
\citep{eerkens2005cultural, o2010innovation}, and suggest that highly
conservative cultural repertoires, such as those posited to precede
behavioral modernity in hominin populations, occur whenever individuals
engage in social learning in small technological design spaces, in the
absence of strong and regular individual innovation.

\subsection{Trait Richness and Knowledge
Depth}\label{trait-richness-and-knowledge-depth}

\begin{figure}[htbp] 
\centering 
\includegraphics[scale=0.4]{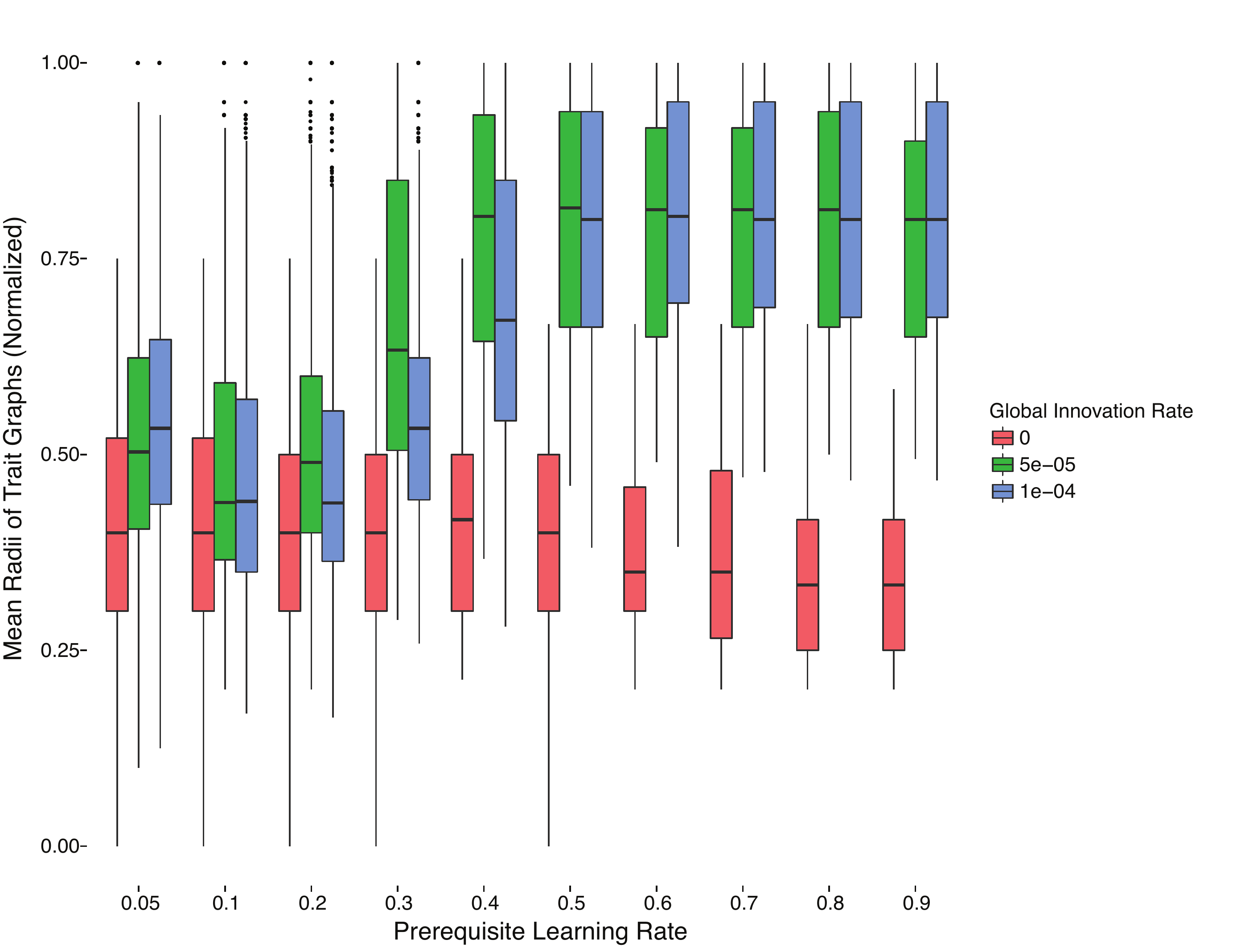} 
\caption{Mean depth of trait sets, by prerequisite learning rate and global innovation rate, for population size 100.} 
\label{img:mean-radius-cultures-100} 
\end{figure}

Cumulative evolution of technology is represented in our model by the
population learning its way \emph{down} the trees which compose the
design space. Possession of traits deeper in the trees represents skills
or information which is more specific, possessing more prerequisites.
Thus, we expect that the depth (or ``radius'', see Figure
\ref{img:final-tree}) of trees would increase with the prerequisite
learning rate, representing a learning environment which is structured
to ensure such acquisition.

Figure \ref{img:mean-radius-cultures-100} gives the \emph{normalized}
mean radius of cultural regions, broken out by the prerequisite learning
rate along the horizontal axis, and each group of 3 boxplots displays
the differing global innovation rates studied. Radii are normalized to
the depth of their design space, to facilitate comparison. The results
indicate that essentially two regimes exist: shorter trees, which do not
grow much beyond their initialized size, and larger trees. The mean
radius has an asymptote just above 0.75, achieved with the prerequisite
learning rate is approximately 0.4 or higher. Further increases do not
seem to matter. Additionally, the difference between the two global
innovation rates is small--what matters most in terms of qualitative
behavior is the presence of global innovation outside the teaching or
learning of prerequisites themselves.

\subsection{Population Size}\label{population-size}

Earlier, we mentioned that population size does not seem to be a primary
factor in explaining the measured diversity in cultural transmission
models, except perhaps in bottleneck situations like the one Henrich
analyzes in Tasmania \citeyearpar{henrich2004}. Instead, population size
may have an interaction effect with other factors, yielding smaller
second-order effects. We examined the effect of population size in the
research reported here, repeating the entire set of simulation runs for
populations of 100, 225, and 400.\footnote{We should note that learning
  rates of 0.8 and 0.9 for population size 400 were cut short due to
  budget constraints, but this does not appear to affect the pattern in
  our dataset.}

\begin{figure}[htbp] 
\centering 
\includegraphics[scale=0.4]{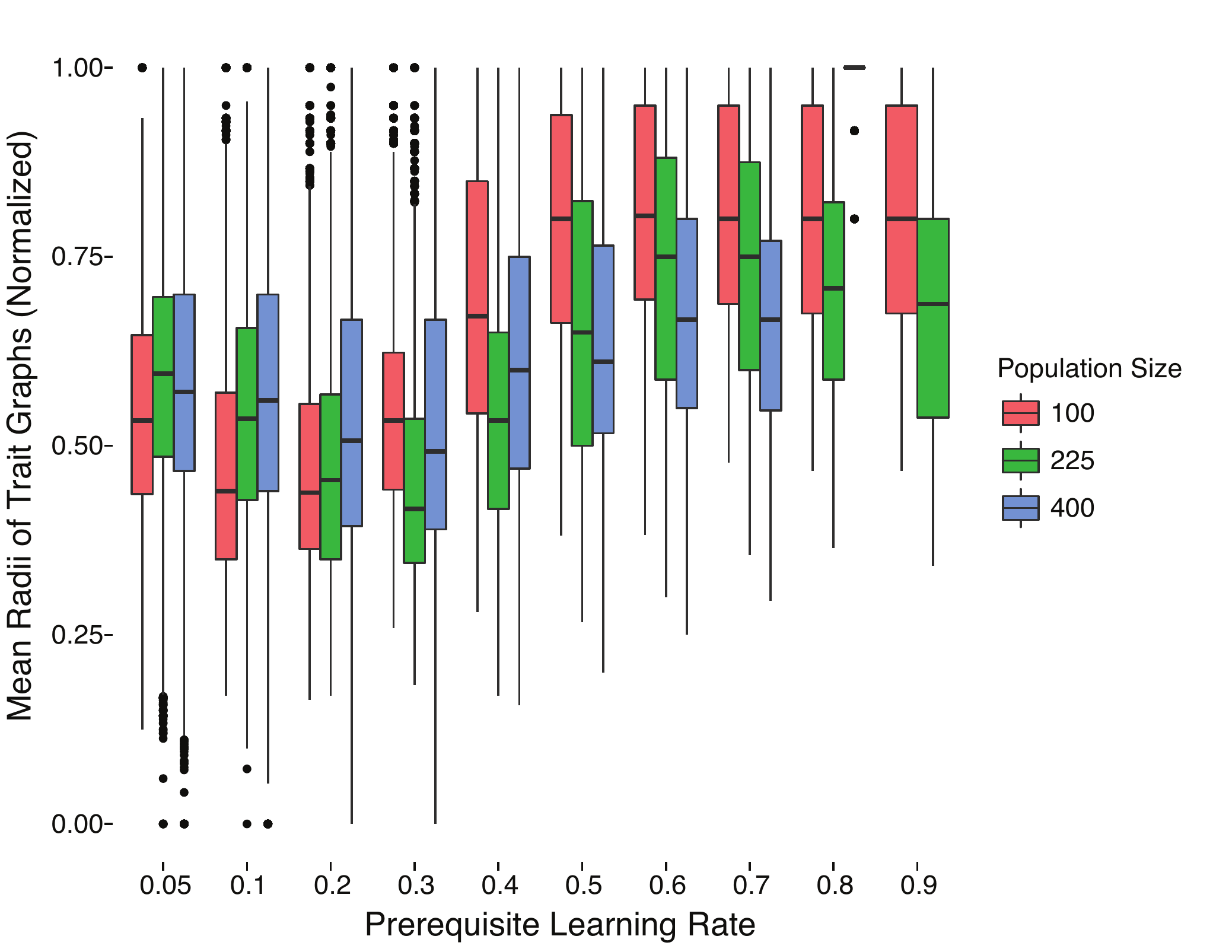} 
\caption{Mean depth of trait sets, by prerequisite learning rate and population sizes of 100, 225 and 400.} 
\label{img:mean-radius-cultures-pop} 
\end{figure}

Figure \ref{img:mean-radius-cultures-pop} displays the relationship
between mean radius (or depth) of the cultural traits in each cultural
sample, as in Figure \ref{img:mean-radius-cultures-100} above, but the
boxplots are instead colored by population size. At least over a range
of group or deme sizes likely to be relevant to Paleolithic archaeology,
population size makes no difference to the qualitative behavior of the
model. There is, however, a very slight decrease in mean radius of trait
sets with larger population size, which is likely a consequence of a
larger population spreading out over the trait space.

\subsection{Trait Tree Symmetries}\label{trait-tree-symmetries}

Finally, we examined the algebraic properties of the trait trees
composing cultural regions, examining both the number of vertex
equivalence classes (orbits) and the size of the automorphism group of
the trait forests. We examined the raw metrics, and versions normalized
by the size of the maximally symmetric forest with the same number of
traits, branching factor, and depth factor. The latter proved difficult
and led to serious overflow problems even with 64 bit arithmetic, so we
focus here on the raw automorphism group size.

\begin{figure}[h] 
\centering 
\includegraphics[scale=0.4]{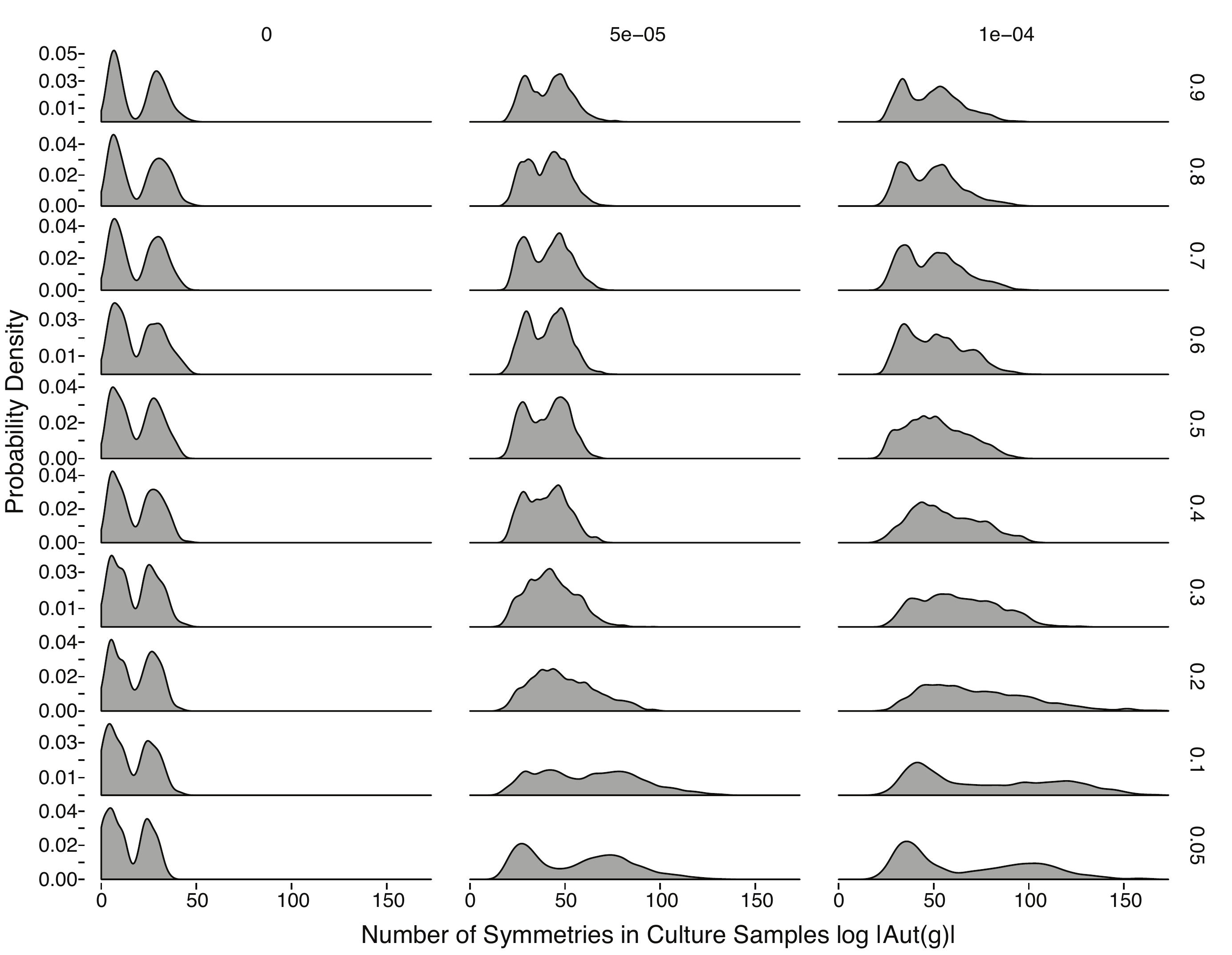} 
\caption{Number of symmetries in trait tree samples, measured as the log of the order of the automorphism group of the trait graphs, broken down by prerequisite learning rate (rows) and global innovation rate (columns).} 
\label{img:autgsize} 
\end{figure}

The logarithm of the automorphism group size does hint at interesting
structure (Figure \ref{img:autgsize}). In the presence of mutation, the
learning of prerequisites narrows the range of variability for the
automorphism group size, and at higher learning rates renders the
distribution multimodal. The modality arises because of the different
combinations of branching factor and depth factor we employed for design
spaces--i.e., some design spaces are ``wide'' and some are ``narrow,''
while also being ``shallow'' or ``deep.'' This gives rise to different
modes in the measured symmetries, but overall the reduction in
variability in symmetry is the most important qualitative effect seen in
our data.

We do not fully understand the ``shapes'' of cultural regions to which
the model appears to converge, but it appears that there is a tendency
for trait graphs to converge towards shapes which have moderate numbers
of symmetries. This graph is on a logarithmic scale, so a peak at 50
along the horizontal axis correponds to a trait graph with approximately
$5 \times 10^{21}$ symmetries. This is a fairly small number, compared
to the original design spaces, which have symmetries ranging from
approximately $10^{41}$ to $10^{6496}$. Thus, the geometry of cultural
traits in our hierarchical design spaces are fairly asymmetric and
represent small and very specific segments of the total design space.

Further analysis of trait graph ``shapes'' is needed to tell whether
there are repeating patterns or graph ``motifs'' which characterize a
social learning model in a graph-structured trait space. The results
here are suggestive of such a phenomenon, but inconclusive given just
the bulk algebraic properties of cultural regions, since the size of the
automorphism group (or the number of orbits) tells only \emph{how many}
symmetries there are, not what types of symmetries exist. The next step
in our analysis of shape is to pursue a geometric decomposition of the
graph following Ben MacArthur and
$\textrm{Rub\'en S\'anchez-Garc\'ia's}$
\citeyearpar{macarthur2008symmetry} work on the symmetries of complex
networks.

\section{Discussion}\label{discussion}

The ``semantic Axelrod'' model described here specifically addresses
social learning of knowledge with ``prerequisite'' structure, and a
learning environment which is tunable from low to high fidelity,
simulating the intensity with which ``teaching'' occurs in addition to
imitative copying. The model displays a characteristic increase in the
cultural repertoires of individuals, as they learn in environments of
higher fidelity. At the individual level, an increase in higher fidelity
learning within structured information environments both creates
path-dependency in what is learned, and increases the chances for
specialization among individuals. Hominin populations in which complex
knowledge is taught systematically along with prerequisites will
accumulate and retain skills and technology faster and to a greater
extent than those groups which rely upon natural pedagogy and imitation
for social learning.

Previous research had established the importance of teaching and
learning environments for cumulative cultural evolution and cultural
diversity
\citep{Aoki2013Determinants-of, Castro201474, Creanza2013Exploring-Cultu, Nakahashi2013Cultural-Evolut}.
Our contribution in this paper is a model capable of connecting the fact
of teaching with the actual structure and content of cultural knowledge.
Such models, we believe, are important in explaining the explosion of
cumulative material culture that accompanies behavioral modernity. The
model described here only makes a start on modeling the additive and
recombinative complexity of real technologies, but it does display
accumulated depth of ``knowledge'' or ``skills,'' as represented by the
radius or depth of trait trees. In combination with realistic models of
technology--such as the production sequences studied by experts on stone
tools--we believe that empirically sufficient models of the evolution of
specific technologies are possible and within reach.

Several areas suggest themselves for future research in structured
information or ``semantic'' cultural transmission models. Some we are
pursuing, others remain open questions and we invite collaboration
towards their solution.

\begin{itemize}
\itemsep1pt\parskip0pt\parsep0pt
\item
  Regional scale cultural differentiation given a metapopulation
  embedding of the basic model.
\item
  Additional trait relations (e.g., class subsumption, functional
  equivalencies).
\item
  Realistic technology models for key artifact classes (e.g., bifaces,
  scrapers, pottery).
\item
  Incorporation of trait fitness in order to study directional change.
\end{itemize}

Models of the class introduced here are ``thicker'' descriptions of how
humans acquire skills and information in real learning environments, and
thus complement existing models which describe the conditions under
which teaching and structured learning might evolve and spread. We
believe models of this type make a needed ``downpayment'' on cultural
transmission models which can substantively incorporate specialties such
as archaeometry, the technnological analysis of lithics and pottery
\citep{tostevin2012seeing}, and studies of how innovation occurs in
various tool classes \citep[e.g.,][]{o2010innovation}. Bringing cultural
transmission modeling together with the details of technologies will be
a crucial component in multifactor evolutionary explanations for the
complex of changes seen in modern \emph{Homo sapiens} and some
Neanderthal populations in the later Paleolithic.

\section{Acknowledgements}\label{acknowledgements}

The authors wish to thank Briggs Buchanan and Mark Collard for the
invitation to participate in the symposium ``Current Research in
Evolutionary Archaeology,'' at the 79th Annual Meeting of the Society
for American Archaeology in Austin, TX. A summary of this research was
presented in that session, and Alex Mesoudi provided valuable comments
on an early post-conference draft. Kenichi Aoki and an anonymous
reviewer provided feedback prior to publication, and although we did not
take all of their suggestions, the comments led to a number of
improvements. Madsen wishes to thank $\textrm{Fr\'ed\'eric}$ Chapoton of
the Institut Camille Jordan for answering a question about the maximal
automorphism group of trees.

\section{Appendices}\label{appendices}

\subsection{Algorithm Description}\label{algorithm-description}

Algorithm \ref{alg:tree-prereq-axelrod} describes the ``semantic''
Axelrod model variant studied in this chapter. Within the algorithm,
there are several functions which find traits with particular
properties. Some, like \textbf{GetTraitUniquetoFocal()}, are fairly
simple set operations but were abbreviated to clarify the notation.

\begin{algorithm}[H]
    \caption{}
    \label{alg:tree-prereq-axelrod}
    \begin{boxedminipage}{\textwidth}
    \begin{algorithmic}[1]
        %\REQUIRE lossrate is the population rate at which traits are randomly lost to drift
        \REQUIRE innovrate is the population rate at which individuals randomly learn a trait
        \REQUIRE learningrate is the probability of learning a missing prerequisite during a learning interaction

        \STATE { $focal \leftarrow$ GetRandomAgent()}
        \STATE { $neighbor \leftarrow$ GetRandomNeighbor(focal)}

        \IF { $focal = neighbor \lor focal \cap neighbor = \varnothing\;\lor  
        neighbor \subsetneq focal $}
        \label{alg:ext-first-if}
            \STATE { exit }
            \COMMENT{ No interaction is possible, move on to next agent }
        \ENDIF

        \STATE { $prob \leftarrow (focal \cup neighbor  - focal \cap neighbor) / focal \cup neighbor$ }

        \IF {RandomUniform() $< prob$}
            \STATE { $differing \leftarrow neighbor \setminus focal$ }
            \STATE { $newtrait \leftarrow$ GetRandomChoice(differing)}

            \IF {hasPrerequisiteForTrait($focal$, $newtrait$) = True}
                \STATE {$replace \leftarrow$ GetTraitUniquetoFocal(focal,neighbor)}
                \STATE { $focal \leftarrow focal \setminus replace$}
                \STATE { $focal \leftarrow focal \cup newtrait$}
            \ELSE
                \IF {RandomUniform() $< learningrate$}
                    \STATE {$prereq \leftarrow$ GetDeepestMissingPrerequisite(newtrait, focal)}
                    \STATE { $focal \leftarrow focal \cup prereq$ }
                \ENDIF
            \ENDIF

        \ENDIF
        %\IF {RandomUniform() $< lossrate$}
        %   \STATE { $focal2 \leftarrow$ GetRandomAgent()}
        %   \STATE { $loss \leftarrow$ GetRandomTrait(focal2)}
        %   \STATE { $focal2 \leftarrow focal2 \setminus loss$}
        %\ENDIF

        \IF {RandomUniform() $< innovrate$}
            \STATE { $focal3 \leftarrow$ GetRandomAgent()}
            \STATE { $innovation \leftarrow$ GetRandomTraitNotInFocal(focal3)}
            \STATE { $focal3 \leftarrow focal3 \cup innovation$}
        \ENDIF

    \end{algorithmic}
    \end{boxedminipage}
\end{algorithm}

\textbf{GetDeepestMissingPrerequisite()} is a procedure which takes the
trait set of an individual, and a trait for which the individual is
known to be missing necessary prerequisites, and returns the ``most
basic'' missing prerequisite for that trait (i.e., closest to the root).
This is done by finding the path which connects the root and desired
trait, and walking its vertices from the root downward, checking to see
if each vertex is part of the individual's trait set. The first trait
not found in the individual's repertoire is returned.

\subsection{Availability of Software and Analysis
Code}\label{availability-of-software-and-analysis-code}

The simulation software used in this chapter is available under an
open-source license at Mark Madsen's GitHub repository
\url{https://github.com/mmadsen/axelrod-ct}. Required libraries and
software are listed in the source archive itself, and include Python 2.7
and the open-source MongoDB database engine to store simulation output.

The codebase consists of a set of library modules which implement the
shared and unique aspects of each model, unit tests to verify the basic
functionality of the code, and scripts which execute each model. The
\textbf{axelrod-ct} repository contains three models:

\begin{itemize}
\item
  An implementation of the original Axelrod model using the
  \textbf{axelrod-ct} libraries.
\item
  A basic model with an ``extensible'' trait space but no relations
  between traits.
\item
  A ``semantic'' Axelrod model with tree-structured trait space
  representing prerequisite relationships between traits.
\end{itemize}

Stepwise extension from the original Axelrod to the semantic models on
the same code library allowed a degree of verification, which is
difficult in a situation where there is no existing mathematical theory
against which to compare the code implementation
\citep{national2012Assessing}.

The analysis and final dataset reported here are available, along with
the source of this paper and associated presentations, in an associated
GitHub repository: \url{https://github.com/mmadsen/madsenlipo2014}.
Statistical analyses of the final dataset were performed in R, rendering
our results reproducible given simulated data from the ``axelrod-ct''
software linked above.


\bibliographystyle{model2-names}
\bibliography{madsenlipo2014-semanticaxelrod}

\begin{thebibliography}{77}
\expandafter\ifx\csname natexlab\endcsname\relax\def\natexlab#1{#1}\fi
\expandafter\ifx\csname url\endcsname\relax
  \def\url#1{\texttt{#1}}\fi
\expandafter\ifx\csname urlprefix\endcsname\relax\def\urlprefix{URL }\fi
\providecommand{\eprint}[2][]{\url{#2}}
\providecommand{\bibinfo}[2]{#2}
\ifx\xfnm\relax \def\xfnm[#1]{\unskip,\space#1}\fi
%Type = Incollection
\bibitem[{Aoki(2013)}]{Aoki2013Determinants-of}
\bibinfo{author}{Aoki, K.}, \bibinfo{year}{2013}.
\newblock \bibinfo{title}{Determinants of cultural evolutionary rates}, in:
  \bibinfo{editor}{Akazawa, T.}, \bibinfo{editor}{Nishiaki, Y.},
  \bibinfo{editor}{Aoki, K.} (Eds.), \bibinfo{booktitle}{Dynamics of Learning
  in Neanderthals and Modern Humans Volume 1}. \bibinfo{publisher}{Springer
  Japan}. Replacement of Neanderthals by Modern Humans Series, pp.
  \bibinfo{pages}{199--210}.
%Type = Article
\bibitem[{Aoki et~al.(2011)Aoki, Lehmann and Feldman}]{aoki2011rates}
\bibinfo{author}{Aoki, K.}, \bibinfo{author}{Lehmann, L.},
  \bibinfo{author}{Feldman, M.W.}, \bibinfo{year}{2011}.
\newblock \bibinfo{title}{Rates of cultural change and patterns of cultural
  accumulation in stochastic models of social transmission}.
\newblock \bibinfo{journal}{Theoretical population biology}
  \bibinfo{volume}{79}, \bibinfo{pages}{192--202}.
%Type = Article
\bibitem[{Axelrod(1997)}]{axelrod1997}
\bibinfo{author}{Axelrod, R.}, \bibinfo{year}{1997}.
\newblock \bibinfo{title}{{The dissemination of culture: A model with local
  convergence and global polarization}}.
\newblock \bibinfo{journal}{Journal of Conflict Resolution}
  \bibinfo{volume}{41}, \bibinfo{pages}{203--226}.
%Type = Article
\bibitem[{Bamforth and Finlay(2008)}]{Bamforth:2008kq}
\bibinfo{author}{Bamforth, D.B.}, \bibinfo{author}{Finlay, N.},
  \bibinfo{year}{2008}.
\newblock \bibinfo{title}{{Introduction: Archaeological Approaches to Lithic
  Production Skill and Craft Learning}}.
\newblock \bibinfo{journal}{Journal of Archaeological Method and Theory}
  \bibinfo{volume}{15}, \bibinfo{pages}{1--27}.
%Type = Article
\bibitem[{Bar-Yosef(2002)}]{bar2002upper}
\bibinfo{author}{Bar-Yosef, O.}, \bibinfo{year}{2002}.
\newblock \bibinfo{title}{The upper paleolithic revolution}.
\newblock \bibinfo{journal}{Annual Review of Anthropology} ,
  \bibinfo{pages}{363--393}.
%Type = Article
\bibitem[{Bleed(2001)}]{bleed2001trees}
\bibinfo{author}{Bleed, P.}, \bibinfo{year}{2001}.
\newblock \bibinfo{title}{Trees or chains, links or branches: conceptual
  alternatives for consideration of stone tool production and other sequential
  activities}.
\newblock \bibinfo{journal}{Journal of Archaeological Method and Theory}
  \bibinfo{volume}{8}, \bibinfo{pages}{101--127}.
%Type = Article
\bibitem[{Bleed(2002)}]{bleed2002obviously}
\bibinfo{author}{Bleed, P.}, \bibinfo{year}{2002}.
\newblock \bibinfo{title}{Obviously sequential, but continuous or staged?
  refits and cognition in three late paleolithic assemblages from japan}.
\newblock \bibinfo{journal}{Journal of Anthropological Archaeology}
  \bibinfo{volume}{21}, \bibinfo{pages}{329--343}.
%Type = Article
\bibitem[{Bleed(2008)}]{Bleed:2008in}
\bibinfo{author}{Bleed, P.}, \bibinfo{year}{2008}.
\newblock \bibinfo{title}{{Skill Matters}}.
\newblock \bibinfo{journal}{Journal of Archaeological Method and Theory}
  \bibinfo{volume}{15}, \bibinfo{pages}{154--166}.
%Type = Article
\bibitem[{Bouzouggar et~al.(2007)Bouzouggar, Barton, Vanhaeren, d'Errico,
  Collcutt, Higham, Hodge, Parfitt, Rhodes, Schwenninger
  et~al.}]{bouzouggar200782}
\bibinfo{author}{Bouzouggar, A.}, \bibinfo{author}{Barton, N.},
  \bibinfo{author}{Vanhaeren, M.}, \bibinfo{author}{d'Errico, F.},
  \bibinfo{author}{Collcutt, S.}, \bibinfo{author}{Higham, T.},
  \bibinfo{author}{Hodge, E.}, \bibinfo{author}{Parfitt, S.},
  \bibinfo{author}{Rhodes, E.}, \bibinfo{author}{Schwenninger, J.L.}, et~al.,
  \bibinfo{year}{2007}.
\newblock \bibinfo{title}{82,000-year-old shell beads from north africa and
  implications for the origins of modern human behavior}.
\newblock \bibinfo{journal}{Proceedings of the National Academy of Sciences}
  \bibinfo{volume}{104}, \bibinfo{pages}{9964--9969}.
%Type = Article
\bibitem[{Castellano et~al.(2009)Castellano, Fortunato and
  Loreto}]{castellano2009statistical}
\bibinfo{author}{Castellano, C.}, \bibinfo{author}{Fortunato, S.},
  \bibinfo{author}{Loreto, V.}, \bibinfo{year}{2009}.
\newblock \bibinfo{title}{Statistical physics of social dynamics}.
\newblock \bibinfo{journal}{Reviews of modern physics} \bibinfo{volume}{81},
  \bibinfo{pages}{591}.
%Type = Article
\bibitem[{Castellano et~al.(2000)Castellano, Marsili and
  Vespignani}]{castellano2000nonequilibrium}
\bibinfo{author}{Castellano, C.}, \bibinfo{author}{Marsili, M.},
  \bibinfo{author}{Vespignani, A.}, \bibinfo{year}{2000}.
\newblock \bibinfo{title}{Nonequilibrium phase transition in a model for social
  influence}.
\newblock \bibinfo{journal}{Physical Review Letters} \bibinfo{volume}{85},
  \bibinfo{pages}{3536}.
%Type = Article
\bibitem[{Castro and Toro(2014)}]{Castro201474}
\bibinfo{author}{Castro, L.}, \bibinfo{author}{Toro, M.A.},
  \bibinfo{year}{2014}.
\newblock \bibinfo{title}{Cumulative cultural evolution: The role of teaching}.
\newblock \bibinfo{journal}{Journal of Theoretical Biology}
  \bibinfo{volume}{347}, \bibinfo{pages}{74 -- 83}.
%Type = Article
\bibitem[{Collard et~al.(2011)Collard, Buchanan, Morin and
  Costopoulos}]{collard2011drives}
\bibinfo{author}{Collard, M.}, \bibinfo{author}{Buchanan, B.},
  \bibinfo{author}{Morin, J.}, \bibinfo{author}{Costopoulos, A.},
  \bibinfo{year}{2011}.
\newblock \bibinfo{title}{What drives the evolution of hunter--gatherer
  subsistence technology? a reanalysis of the risk hypothesis with data from
  the pacific northwest}.
\newblock \bibinfo{journal}{Philosophical Transactions of the Royal Society B:
  Biological Sciences} \bibinfo{volume}{366}, \bibinfo{pages}{1129--1138}.
%Type = Article
\bibitem[{Collard et~al.(2013a)Collard, Buchanan and
  OBrien}]{collard2013population}
\bibinfo{author}{Collard, M.}, \bibinfo{author}{Buchanan, B.},
  \bibinfo{author}{OBrien, M.J.}, \bibinfo{year}{2013}a.
\newblock \bibinfo{title}{Population size as an explanation for patterns in the
  paleolithic archaeological record}.
\newblock \bibinfo{journal}{Current Anthropology} \bibinfo{volume}{54},
  \bibinfo{pages}{S388--S396}.
%Type = Article
\bibitem[{Collard et~al.(2013b)Collard, Buchanan, O'Brien and
  Scholnick}]{collard2013risk}
\bibinfo{author}{Collard, M.}, \bibinfo{author}{Buchanan, B.},
  \bibinfo{author}{O'Brien, M.J.}, \bibinfo{author}{Scholnick, J.},
  \bibinfo{year}{2013}b.
\newblock \bibinfo{title}{Risk, mobility or population size? drivers of
  technological richness among contact-period western north american
  hunter--gatherers}.
\newblock \bibinfo{journal}{Philosophical Transactions of the Royal Society B:
  Biological Sciences} \bibinfo{volume}{368}, \bibinfo{pages}{20120412}.
%Type = Article
\bibitem[{Collard et~al.(2013c)Collard, Ruttle, Buchanan and
  OBrien}]{collard2013plos}
\bibinfo{author}{Collard, M.}, \bibinfo{author}{Ruttle, A.},
  \bibinfo{author}{Buchanan, B.}, \bibinfo{author}{OBrien, M.J.},
  \bibinfo{year}{2013}c.
\newblock \bibinfo{title}{Population size and cultural evolution in
  nonindustrial food-producing societies}.
\newblock \bibinfo{journal}{PloS one} \bibinfo{volume}{8},
  \bibinfo{pages}{e72628}.
%Type = Book
\bibitem[{{Committee on Mathematical Foundations of Verification Validation and
  Uncertainty Quantification, National Research
  Council}(2012)}]{national2012Assessing}
\bibinfo{author}{{Committee on Mathematical Foundations of Verification
  Validation and Uncertainty Quantification, National Research Council}},
  \bibinfo{year}{2012}.
\newblock \bibinfo{title}{Assessing the Reliability of Complex Models:
  Mathematical and Statistical Foundations of Verification, Validation, and
  Uncertainty Quantification}.
\newblock \bibinfo{publisher}{The National Academies Press}.
%Type = Incollection
\bibitem[{Creanza et~al.(2013)Creanza, Fogarty and
  Feldman}]{Creanza2013Exploring-Cultu}
\bibinfo{author}{Creanza, N.}, \bibinfo{author}{Fogarty, L.},
  \bibinfo{author}{Feldman, M.}, \bibinfo{year}{2013}.
\newblock \bibinfo{title}{Exploring cultural niche construction from the
  paleolithic to modern hunter-gatherers}, in: \bibinfo{editor}{Akazawa, T.},
  \bibinfo{editor}{Nishiaki, Y.}, \bibinfo{editor}{Aoki, K.} (Eds.),
  \bibinfo{booktitle}{Dynamics of Learning in Neanderthals and Modern Humans
  Volume 1}. \bibinfo{publisher}{Springer Japan}. Replacement of Neanderthals
  by Modern Humans Series, pp. \bibinfo{pages}{211--228}.
%Type = Article
\bibitem[{Csibra and Gergely(2011)}]{Csibra:2011dx}
\bibinfo{author}{Csibra, G.}, \bibinfo{author}{Gergely, G.},
  \bibinfo{year}{2011}.
\newblock \bibinfo{title}{{Natural pedagogy as evolutionary adaptation.}}
\newblock \bibinfo{journal}{Philosophical Transactions of the Royal Society B:
  Biological Sciences} \bibinfo{volume}{366}, \bibinfo{pages}{1149--1157}.
%Type = Article
\bibitem[{De~Sanctis and Galla(2009)}]{de2009effects}
\bibinfo{author}{De~Sanctis, L.}, \bibinfo{author}{Galla, T.},
  \bibinfo{year}{2009}.
\newblock \bibinfo{title}{Effects of noise and confidence thresholds in nominal
  and metric axelrod dynamics of social influence}.
\newblock \bibinfo{journal}{Physical Review E} \bibinfo{volume}{79},
  \bibinfo{pages}{046108}.
%Type = Article
\bibitem[{Derex et~al.(2013)Derex, Beugin, Godelle and
  Raymond}]{derex2013experimental}
\bibinfo{author}{Derex, M.}, \bibinfo{author}{Beugin, M.P.},
  \bibinfo{author}{Godelle, B.}, \bibinfo{author}{Raymond, M.},
  \bibinfo{year}{2013}.
\newblock \bibinfo{title}{Experimental evidence for the influence of group size
  on cultural complexity}.
\newblock \bibinfo{journal}{Nature} \bibinfo{volume}{503},
  \bibinfo{pages}{389--391}.
%Type = Article
\bibitem[{d'Errico and Henshilwood(2007)}]{d2007additional}
\bibinfo{author}{d'Errico, F.}, \bibinfo{author}{Henshilwood, C.S.},
  \bibinfo{year}{2007}.
\newblock \bibinfo{title}{Additional evidence for bone technology in the
  southern african middle stone age}.
\newblock \bibinfo{journal}{Journal of Human Evolution} \bibinfo{volume}{52},
  \bibinfo{pages}{142--163}.
%Type = Article
\bibitem[{d'Errico and Stringer(2011)}]{d2011evolution}
\bibinfo{author}{d'Errico, F.}, \bibinfo{author}{Stringer, C.B.},
  \bibinfo{year}{2011}.
\newblock \bibinfo{title}{Evolution, revolution or saltation scenario for the
  emergence of modern cultures?}
\newblock \bibinfo{journal}{Philosophical Transactions of the Royal Society B:
  Biological Sciences} \bibinfo{volume}{366}, \bibinfo{pages}{1060--1069}.
%Type = Book
\bibitem[{Diestel(2010)}]{diestel2010graph}
\bibinfo{author}{Diestel, R.}, \bibinfo{year}{2010}.
\newblock \bibinfo{title}{Graph Theory}.
\newblock Graduate Texts in Mathematics Vol. 173,
  \bibinfo{publisher}{Springer-Verlag, Heidelberg}. \bibinfo{edition}{4th}
  edition.
%Type = Book
\bibitem[{Dunnell(1971)}]{Dunnell1971}
\bibinfo{author}{Dunnell, R.C.}, \bibinfo{year}{1971}.
\newblock \bibinfo{title}{Systematics in prehistory}.
\newblock \bibinfo{publisher}{Free Press}, \bibinfo{address}{New York}.
%Type = Article
\bibitem[{Eerkens and Lipo(2005)}]{eerkens2005cultural}
\bibinfo{author}{Eerkens, J.}, \bibinfo{author}{Lipo, C.},
  \bibinfo{year}{2005}.
\newblock \bibinfo{title}{Cultural transmission, copying errors, and the
  generation of variation in material culture and the archaeological record}.
\newblock \bibinfo{journal}{Journal of Anthropological Archaeology}
  \bibinfo{volume}{24}, \bibinfo{pages}{316--334}.
%Type = Book
\bibitem[{Ewens(2004)}]{Ewens2004}
\bibinfo{author}{Ewens, W.J.}, \bibinfo{year}{2004}.
\newblock \bibinfo{title}{Mathematical Population Genetics, Volume 1:
  Theoretical Introduction}.
\newblock \bibinfo{publisher}{New York, Springer}. \bibinfo{edition}{2nd}
  edition.
%Type = Article
\bibitem[{Ferguson(2008)}]{Ferguson:2008ce}
\bibinfo{author}{Ferguson, J.R.}, \bibinfo{year}{2008}.
\newblock \bibinfo{title}{{The When, Where, and How of Novices in Craft
  Production}}.
\newblock \bibinfo{journal}{Journal of Archaeological Method and Theory}
  \bibinfo{volume}{15}, \bibinfo{pages}{51--67}.
%Type = Article
\bibitem[{Flache and Macy(2006)}]{flache2006sustains}
\bibinfo{author}{Flache, A.}, \bibinfo{author}{Macy, M.W.},
  \bibinfo{year}{2006}.
\newblock \bibinfo{title}{What sustains cultural diversity and what undermines
  it? axelrod and beyond}.
\newblock \bibinfo{journal}{arXiv preprint physics/0604201} .
%Type = Article
\bibitem[{Fogarty et~al.(2011)Fogarty, Strimling and Laland}]{Fogarty:2011gv}
\bibinfo{author}{Fogarty, L.}, \bibinfo{author}{Strimling, P.},
  \bibinfo{author}{Laland, K.N.}, \bibinfo{year}{2011}.
\newblock \bibinfo{title}{{The evolution of teaching.}}
\newblock \bibinfo{journal}{Evolution} \bibinfo{volume}{65},
  \bibinfo{pages}{2760--2770}.
%Type = Book
\bibitem[{Godsil and Royle(2001)}]{godsil2001algebraic}
\bibinfo{author}{Godsil, C.D.}, \bibinfo{author}{Royle, G.},
  \bibinfo{year}{2001}.
\newblock \bibinfo{title}{Algebraic graph theory}. volume~\bibinfo{volume}{8}.
\newblock \bibinfo{publisher}{Springer New York}.
%Type = Article
\bibitem[{Gonzalez-Avella et~al.(2007a)Gonzalez-Avella, Cosenza and
  Klemm}]{GonzalezAvella:2007p6912}
\bibinfo{author}{Gonzalez-Avella, J.}, \bibinfo{author}{Cosenza, M.},
  \bibinfo{author}{Klemm, K.}, \bibinfo{year}{2007}a.
\newblock \bibinfo{title}{{Information feedback and mass media effects in
  cultural dynamics}}.
\newblock \bibinfo{journal}{Journal of Artificial Societies and Social
  Simulation} .
%Type = Article
\bibitem[{Gonzalez-Avella et~al.(2007b)Gonzalez-Avella, Eguiluz and
  San~Miguel}]{GonzalezAvella:2007p6910}
\bibinfo{author}{Gonzalez-Avella, J.}, \bibinfo{author}{Eguiluz, V.},
  \bibinfo{author}{San~Miguel, M.}, \bibinfo{year}{2007}b.
\newblock \bibinfo{title}{{Homophily, Cultural Drift, and the Co-Evolution of
  Cultural Groups}}.
\newblock \bibinfo{journal}{Journal of Conflict Resolution} .
%Type = Article
\bibitem[{Gonz{\'a}lez-Avella et~al.(2005)Gonz{\'a}lez-Avella, Cosenza and
  Tucci}]{gonzalez2005nonequilibrium}
\bibinfo{author}{Gonz{\'a}lez-Avella, J.C.}, \bibinfo{author}{Cosenza, M.G.},
  \bibinfo{author}{Tucci, K.}, \bibinfo{year}{2005}.
\newblock \bibinfo{title}{Nonequilibrium transition induced by mass media in a
  model for social influence}.
\newblock \bibinfo{journal}{Physical Review E} \bibinfo{volume}{72},
  \bibinfo{pages}{065102}.
%Type = Article
\bibitem[{Gonz{\'a}lez-Avella et~al.(2006)Gonz{\'a}lez-Avella, Egu{\'\i}luz,
  Cosenza, Klemm, Herrera and San~Miguel}]{gonzalez2006local}
\bibinfo{author}{Gonz{\'a}lez-Avella, J.C.}, \bibinfo{author}{Egu{\'\i}luz,
  V.M.}, \bibinfo{author}{Cosenza, M.G.}, \bibinfo{author}{Klemm, K.},
  \bibinfo{author}{Herrera, J.}, \bibinfo{author}{San~Miguel, M.},
  \bibinfo{year}{2006}.
\newblock \bibinfo{title}{Local versus global interactions in nonequilibrium
  transitions: A model of social dynamics}.
\newblock \bibinfo{journal}{Physical Review E} \bibinfo{volume}{73},
  \bibinfo{pages}{046119}.
%Type = Article
\bibitem[{Henrich(2004)}]{henrich2004}
\bibinfo{author}{Henrich, J.}, \bibinfo{year}{2004}.
\newblock \bibinfo{title}{Demography and cultural evolution: how adaptive
  cultural processes can produce maladaptive losses: the tasmanian case}.
\newblock \bibinfo{journal}{American Antiquity} \bibinfo{volume}{69},
  \bibinfo{pages}{197--214}.
%Type = Article
\bibitem[{H{\"o}gberg(2008)}]{Hogberg:2008fj}
\bibinfo{author}{H{\"o}gberg, A.}, \bibinfo{year}{2008}.
\newblock \bibinfo{title}{{Playing with Flint: Tracing a Child's Imitation of
  Adult Work in a Lithic Assemblage}}.
\newblock \bibinfo{journal}{Journal of Archaeological Method and Theory}
  \bibinfo{volume}{15}, \bibinfo{pages}{112--131}.
%Type = Article
\bibitem[{Kempe and Mesoudi(2014)}]{kempe2014experimental}
\bibinfo{author}{Kempe, M.}, \bibinfo{author}{Mesoudi, A.},
  \bibinfo{year}{2014}.
\newblock \bibinfo{title}{An experimental demonstration of the effect of group
  size on cultural accumulation}.
\newblock \bibinfo{journal}{Evolution and Human Behavior} .
%Type = Book
\bibitem[{Klein(2009)}]{klein2009human}
\bibinfo{author}{Klein, R.G.}, \bibinfo{year}{2009}.
\newblock \bibinfo{title}{The human career: human biological and cultural
  origins}.
\newblock \bibinfo{publisher}{University of Chicago Press}.
%Type = Article
\bibitem[{Klemm et~al.(2003a)Klemm, Egu{\'\i}luz, Toral and
  Miguel}]{Klemm:2003p7031}
\bibinfo{author}{Klemm, K.}, \bibinfo{author}{Egu{\'\i}luz, V.},
  \bibinfo{author}{Toral, R.}, \bibinfo{author}{Miguel, M.},
  \bibinfo{year}{2003}a.
\newblock \bibinfo{title}{{Global culture: A noise-induced transition in finite
  systems}}.
\newblock \bibinfo{journal}{Physical Review E} .
%Type = Article
\bibitem[{Klemm et~al.(2003b)Klemm, Egu{\'\i}luz, Toral and
  San~Miguel}]{Klemm:2003p7112}
\bibinfo{author}{Klemm, K.}, \bibinfo{author}{Egu{\'\i}luz, V.},
  \bibinfo{author}{Toral, R.}, \bibinfo{author}{San~Miguel, M.},
  \bibinfo{year}{2003}b.
\newblock \bibinfo{title}{{Nonequilibrium transitions in complex networks: A
  model of social interaction}}.
\newblock \bibinfo{journal}{Physical Review E} .
%Type = Article
\bibitem[{Klemm et~al.(2005)Klemm, Eguı́luz, Toral and Miguel}]{Klemm:2005tb}
\bibinfo{author}{Klemm, K.}, \bibinfo{author}{Eguı́luz, V.},
  \bibinfo{author}{Toral, R.}, \bibinfo{author}{Miguel, M.},
  \bibinfo{year}{2005}.
\newblock \bibinfo{title}{{Globalization, polarization and cultural drift}}.
\newblock \bibinfo{journal}{Journal of Economic Dynamics and Control} .
%Type = Incollection
\bibitem[{Kuhn(2013)}]{Kuhn2013Cultural-Transm}
\bibinfo{author}{Kuhn, S.}, \bibinfo{year}{2013}.
\newblock \bibinfo{title}{Cultural transmission, institutional continuity and
  the persistence of the mousterian}, in: \bibinfo{editor}{Akazawa, T.},
  \bibinfo{editor}{Nishiaki, Y.}, \bibinfo{editor}{Aoki, K.} (Eds.),
  \bibinfo{booktitle}{Dynamics of Learning in Neanderthals and Modern Humans
  Volume 1}. \bibinfo{publisher}{Springer Japan}. Replacement of Neanderthals
  by Modern Humans Series, pp. \bibinfo{pages}{105--113}.
%Type = Article
\bibitem[{Lanchier(2012)}]{Lanchier:2012ur}
\bibinfo{author}{Lanchier, N.}, \bibinfo{year}{2012}.
\newblock \bibinfo{title}{{The Axelrod model for the dissemination of culture
  revisited}}.
\newblock \bibinfo{journal}{The Annals of Applied Probability}
  \bibinfo{volume}{22}, \bibinfo{pages}{860--880}.
%Type = Article
\bibitem[{Lanchier et~al.(2010)Lanchier, Deijfen, H{\"a}ggstr{\"o}m and
  Connor}]{Lanchier:2010p16999}
\bibinfo{author}{Lanchier, N.}, \bibinfo{author}{Deijfen, M.},
  \bibinfo{author}{H{\"a}ggstr{\"o}m, O.}, \bibinfo{author}{Connor, S.},
  \bibinfo{year}{2010}.
\newblock \bibinfo{title}{{Opinion dynamics with confidence threshold: an
  alternative to the Axelrod model}}.
\newblock \bibinfo{journal}{Alea} .
%Type = Article
\bibitem[{MacArthur et~al.(2008)MacArthur, S{\'a}nchez-Garc{\'\i}a and
  Anderson}]{macarthur2008symmetry}
\bibinfo{author}{MacArthur, B.D.}, \bibinfo{author}{S{\'a}nchez-Garc{\'\i}a,
  R.J.}, \bibinfo{author}{Anderson, J.W.}, \bibinfo{year}{2008}.
\newblock \bibinfo{title}{Symmetry in complex networks}.
\newblock \bibinfo{journal}{Discrete Applied Mathematics}
  \bibinfo{volume}{156}, \bibinfo{pages}{3525--3531}.
%Type = Article
\bibitem[{McBrearty(2007)}]{mcbrearty2007down}
\bibinfo{author}{McBrearty, S.}, \bibinfo{year}{2007}.
\newblock \bibinfo{title}{Down with the revolution}.
\newblock \bibinfo{journal}{Rethinking the human revolution. Cambridge:
  MacDonald Institute for Archaeological Research Monographs} ,
  \bibinfo{pages}{133--152}.
%Type = Article
\bibitem[{McBrearty and Brooks(2000)}]{mcbrearty2000revolution}
\bibinfo{author}{McBrearty, S.}, \bibinfo{author}{Brooks, A.S.},
  \bibinfo{year}{2000}.
\newblock \bibinfo{title}{The revolution that wasn't: a new interpretation of
  the origin of modern human behavior}.
\newblock \bibinfo{journal}{Journal of human evolution} \bibinfo{volume}{39},
  \bibinfo{pages}{453--563}.
%Type = Article
\bibitem[{McKay and Piperno(2014)}]{McKay201494}
\bibinfo{author}{McKay, B.D.}, \bibinfo{author}{Piperno, A.},
  \bibinfo{year}{2014}.
\newblock \bibinfo{title}{Practical graph isomorphism, \{II\}}.
\newblock \bibinfo{journal}{Journal of Symbolic Computation}
  \bibinfo{volume}{60}, \bibinfo{pages}{94 -- 112}.
%Type = Article
\bibitem[{Mesoudi and O'Brien(2008)}]{Mesoudi2008a}
\bibinfo{author}{Mesoudi, A.}, \bibinfo{author}{O'Brien, M.J.},
  \bibinfo{year}{2008}.
\newblock \bibinfo{title}{The learning and transmission of hierarchical
  cultural recipes}.
\newblock \bibinfo{journal}{Biological Theory} \bibinfo{volume}{3},
  \bibinfo{pages}{63--72}.
%Type = Incollection
\bibitem[{Moore(2010)}]{moore2010grammars}
\bibinfo{author}{Moore, M.W.}, \bibinfo{year}{2010}.
\newblock \bibinfo{title}{``grammars of action'' and stone flaking design
  space}, in: \bibinfo{editor}{Nowell, A.}, \bibinfo{editor}{Davidson, I.}
  (Eds.), \bibinfo{booktitle}{Stone tools and the evolution of human
  cognition}. \bibinfo{publisher}{University of Colorado Press, Boulder}, pp.
  \bibinfo{pages}{13--43}.
%Type = Inproceedings
\bibitem[{Moran(1958)}]{moran1958random}
\bibinfo{author}{Moran, P.}, \bibinfo{year}{1958}.
\newblock \bibinfo{title}{Random processes in genetics}, in:
  \bibinfo{booktitle}{Mathematical Proceedings of the Cambridge Philosophical
  Society}, \bibinfo{organization}{Cambridge Univ Press}. pp.
  \bibinfo{pages}{60--71}.
%Type = Article
\bibitem[{Moran et~al.(1962)}]{moran1962statistical}
\bibinfo{author}{Moran, P.}, et~al., \bibinfo{year}{1962}.
\newblock \bibinfo{title}{The statistical processes of evolutionary theory.}
\newblock \bibinfo{journal}{The statistical processes of evolutionary theory.}
  .
%Type = Article
\bibitem[{Muthukrishna et~al.(2014)Muthukrishna, Shulman, Vasilescu and
  Henrich}]{muthukrishna2014sociality}
\bibinfo{author}{Muthukrishna, M.}, \bibinfo{author}{Shulman, B.W.},
  \bibinfo{author}{Vasilescu, V.}, \bibinfo{author}{Henrich, J.},
  \bibinfo{year}{2014}.
\newblock \bibinfo{title}{Sociality influences cultural complexity}.
\newblock \bibinfo{journal}{Proceedings of the Royal Society B: Biological
  Sciences} \bibinfo{volume}{281}, \bibinfo{pages}{20132511}.
%Type = Incollection
\bibitem[{Nakahashi(2013)}]{Nakahashi2013Cultural-Evolut}
\bibinfo{author}{Nakahashi, W.}, \bibinfo{year}{2013}.
\newblock \bibinfo{title}{Cultural evolution and learning strategies in
  hominids}, in: \bibinfo{editor}{Akazawa, T.}, \bibinfo{editor}{Nishiaki, Y.},
  \bibinfo{editor}{Aoki, K.} (Eds.), \bibinfo{booktitle}{Dynamics of Learning
  in Neanderthals and Modern Humans Volume 1}. \bibinfo{publisher}{Springer
  Japan}. Replacement of Neanderthals by Modern Humans Series, pp.
  \bibinfo{pages}{245--254}.
%Type = Article
\bibitem[{Neff(1992)}]{neff1992ceramics}
\bibinfo{author}{Neff, H.}, \bibinfo{year}{1992}.
\newblock \bibinfo{title}{Ceramics and evolution}.
\newblock \bibinfo{journal}{Archaeological Method and Theory}
  \bibinfo{volume}{4}, \bibinfo{pages}{141--193}.
%Type = Inproceedings
\bibitem[{Nickel et~al.(2011)Nickel, Tresp and Kriegel}]{ICML2011Nickel_438}
\bibinfo{author}{Nickel, M.}, \bibinfo{author}{Tresp, V.},
  \bibinfo{author}{Kriegel, H.P.}, \bibinfo{year}{2011}.
\newblock \bibinfo{title}{A three-way model for collective learning on
  multi-relational data}, in: \bibinfo{editor}{Getoor, L.},
  \bibinfo{editor}{Scheffer, T.} (Eds.), \bibinfo{booktitle}{Proceedings of the
  28th International Conference on Machine Learning (ICML-11)},
  \bibinfo{publisher}{ACM}, \bibinfo{address}{New York, NY, USA}. pp.
  \bibinfo{pages}{809--816}.
%Type = Incollection
\bibitem[{Nishiaki et~al.(2013)Nishiaki, Aoki and
  Akazawa}]{Nishiaki2013Introduction}
\bibinfo{author}{Nishiaki, Y.}, \bibinfo{author}{Aoki, K.},
  \bibinfo{author}{Akazawa, T.}, \bibinfo{year}{2013}.
\newblock \bibinfo{title}{Introduction}, in: \bibinfo{editor}{Akazawa, T.},
  \bibinfo{editor}{Nishiaki, Y.}, \bibinfo{editor}{Aoki, K.} (Eds.),
  \bibinfo{booktitle}{Dynamics of Learning in Neanderthals and Modern Humans
  Volume 1}. \bibinfo{publisher}{Springer Japan}. Replacement of Neanderthals
  by Modern Humans Series, pp. \bibinfo{pages}{1--3}.
%Type = Article
\bibitem[{O'Brien et~al.(2010)O'Brien, Lyman, Mesoudi and
  VanPool}]{o2010cultural}
\bibinfo{author}{O'Brien, M.}, \bibinfo{author}{Lyman, R.},
  \bibinfo{author}{Mesoudi, A.}, \bibinfo{author}{VanPool, T.},
  \bibinfo{year}{2010}.
\newblock \bibinfo{title}{Cultural traits as units of analysis}.
\newblock \bibinfo{journal}{Philosophical Transactions of the Royal Society B:
  Biological Sciences} \bibinfo{volume}{365}, \bibinfo{pages}{3797--3806}.
%Type = Book
\bibitem[{O'Brien and Shennan(2010)}]{o2010innovation}
\bibinfo{author}{O'Brien, M.J.}, \bibinfo{author}{Shennan, S.},
  \bibinfo{year}{2010}.
\newblock \bibinfo{title}{Innovation in cultural systems: Contributions from
  evolutionary anthropology}.
\newblock \bibinfo{publisher}{MIT Press}.
%Type = Article
\bibitem[{Otter(1948)}]{otter1948number}
\bibinfo{author}{Otter, R.}, \bibinfo{year}{1948}.
\newblock \bibinfo{title}{The number of trees}.
\newblock \bibinfo{journal}{The Annals of Mathematics} \bibinfo{volume}{49},
  \bibinfo{pages}{583--599}.
%Type = Article
\bibitem[{Premo(2012)}]{premo2012local}
\bibinfo{author}{Premo, L.}, \bibinfo{year}{2012}.
\newblock \bibinfo{title}{Local extinctions, connectedness, and cultural
  evolution in structured populations}.
\newblock \bibinfo{journal}{Advances in Complex Systems} \bibinfo{volume}{15}.
%Type = Misc
\bibitem[{Rotman(1995)}]{rotman1995introduction}
\bibinfo{author}{Rotman, J.J.}, \bibinfo{year}{1995}.
\newblock \bibinfo{title}{An introduction to the theory of groups, volume 148
  of graduate texts in mathematics}.
%Type = Article
\bibitem[{Schank and Abelson(1977)}]{schank1977scripts}
\bibinfo{author}{Schank, R.C.}, \bibinfo{author}{Abelson, R.P.},
  \bibinfo{year}{1977}.
\newblock \bibinfo{title}{Scripts, plans, goals, and understanding: An inquiry
  into human knowledge structures (artificial intelligence series)} .
%Type = Article
\bibitem[{Schiffer and Skibo(1987)}]{schiffer1987theory}
\bibinfo{author}{Schiffer, M.B.}, \bibinfo{author}{Skibo, J.M.},
  \bibinfo{year}{1987}.
\newblock \bibinfo{title}{Theory and experiment in the study of technological
  change}.
\newblock \bibinfo{journal}{Current Anthropology} \bibinfo{volume}{28},
  \bibinfo{pages}{595--622}.
%Type = Article
\bibitem[{Shennan(2000)}]{shennan2000population}
\bibinfo{author}{Shennan, S.}, \bibinfo{year}{2000}.
\newblock \bibinfo{title}{Population, culture history, and the dynamics of
  culture change1}.
\newblock \bibinfo{journal}{Current Anthropology} \bibinfo{volume}{41},
  \bibinfo{pages}{811--835}.
%Type = Article
\bibitem[{Shennan(2001)}]{shennan2001demography}
\bibinfo{author}{Shennan, S.}, \bibinfo{year}{2001}.
\newblock \bibinfo{title}{Demography and cultural innovation: a model and its
  implications for the emergence of modern human culture}.
\newblock \bibinfo{journal}{Cambridge Archaeological Journal}
  \bibinfo{volume}{11}, \bibinfo{pages}{5--16}.
%Type = Article
\bibitem[{Smilkov and Kocarev(2012)}]{smilkov2012influence}
\bibinfo{author}{Smilkov, D.}, \bibinfo{author}{Kocarev, L.},
  \bibinfo{year}{2012}.
\newblock \bibinfo{title}{Influence of the network topology on epidemic
  spreading}.
\newblock \bibinfo{journal}{Physical Review E} \bibinfo{volume}{85},
  \bibinfo{pages}{016114}.
%Type = Book
\bibitem[{Sterelny(2012)}]{sterelny2012evolved}
\bibinfo{author}{Sterelny, K.}, \bibinfo{year}{2012}.
\newblock \bibinfo{title}{The evolved apprentice}.
\newblock \bibinfo{publisher}{MIT Press}.
%Type = Article
\bibitem[{Stout(2002)}]{stout2002skill}
\bibinfo{author}{Stout, D.}, \bibinfo{year}{2002}.
\newblock \bibinfo{title}{Skill and cognition in stone tool production: An
  ethnographic case study from irian jaya 1}.
\newblock \bibinfo{journal}{Current Anthropology} \bibinfo{volume}{43},
  \bibinfo{pages}{693--722}.
%Type = Article
\bibitem[{Stout(2011)}]{stout2011stone}
\bibinfo{author}{Stout, D.}, \bibinfo{year}{2011}.
\newblock \bibinfo{title}{Stone toolmaking and the evolution of human culture
  and cognition}.
\newblock \bibinfo{journal}{Philosophical Transactions of the Royal Society B:
  Biological Sciences} \bibinfo{volume}{366}, \bibinfo{pages}{1050--1059}.
%Type = Article
\bibitem[{Straus(2005)}]{guy2005mosaic}
\bibinfo{author}{Straus, L.G.}, \bibinfo{year}{2005}.
\newblock \bibinfo{title}{A mosaic of change: the middle--upper paleolithic
  transition as viewed from new mexico and iberia}.
\newblock \bibinfo{journal}{Quaternary international} \bibinfo{volume}{137},
  \bibinfo{pages}{47--67}.
%Type = Incollection
\bibitem[{Terashima(2013)}]{Terashima2013The-Evolutionar}
\bibinfo{author}{Terashima, H.}, \bibinfo{year}{2013}.
\newblock \bibinfo{title}{The evolutionary development of learning and teaching
  strategies in human societies}, in: \bibinfo{editor}{Akazawa, T.},
  \bibinfo{editor}{Nishiaki, Y.}, \bibinfo{editor}{Aoki, K.} (Eds.),
  \bibinfo{booktitle}{Dynamics of Learning in Neanderthals and Modern Humans
  Volume 1}. \bibinfo{publisher}{Springer Japan}. Replacement of Neanderthals
  by Modern Humans Series, pp. \bibinfo{pages}{141--150}.
%Type = Book
\bibitem[{Tostevin(2012)}]{tostevin2012seeing}
\bibinfo{author}{Tostevin, G.B.}, \bibinfo{year}{2012}.
\newblock \bibinfo{title}{Seeing lithics: a middle-range theory for testing for
  cultural transmission in the pleistocene}.
\newblock \bibinfo{publisher}{Oxford: Oxbow Books}.
%Type = Article
\bibitem[{Villa and Roebroeks(2014)}]{Villa:2014kl}
\bibinfo{author}{Villa, P.}, \bibinfo{author}{Roebroeks, W.},
  \bibinfo{year}{2014}.
\newblock \bibinfo{title}{{Neandertal Demise: An Archaeological Analysis of the
  Modern Human Superiority Complex}}.
\newblock \bibinfo{journal}{PLoS ONE} \bibinfo{volume}{9},
  \bibinfo{pages}{e96424}.
%Type = Book
\bibitem[{Wimsatt(2007)}]{wimsatt2007re}
\bibinfo{author}{Wimsatt, W.C.}, \bibinfo{year}{2007}.
\newblock \bibinfo{title}{Re-engineering philosophy for limited beings:
  Piecewise approximations to reality}.
\newblock \bibinfo{publisher}{Harvard University Press}.
%Type = Article
\bibitem[{Wimsatt and Griesemer(2007)}]{wimsatt2007reproducing}
\bibinfo{author}{Wimsatt, W.C.}, \bibinfo{author}{Griesemer, J.R.},
  \bibinfo{year}{2007}.
\newblock \bibinfo{title}{Reproducing entrenchments to scaffold culture: The
  central role of development in cultural evolution}.
\newblock \bibinfo{journal}{Integrating evolution and development: From theory
  to practice} , \bibinfo{pages}{227--323}.

\end{thebibliography}

\end{document}